\documentclass[twocolumn,usenatbib]{aastex63}

\usepackage{amsmath}
\usepackage[T1]{fontenc}

\received{XXX}
\revised{YYY}
\accepted{ZZZ}
\submitjournal{ApJ}

\shorttitle{Characterizing the dust content in Sz\,91}
\shortauthors{Mauc\'o et al.}
\graphicspath{{./}{figures/}}

\begin{document}

\title{Characterization of the dust content in the ring around Sz\,91: indications for planetesimal formation?}

\email{karina.mauco@uv.cl}

\author[0000-0001-8284-4343]{Karina Mauc\'o}
\affiliation{Instituto  de  F\'isica  y  Astronom\'ia,
Facultad  de  Ciencias,  Universidad de Valpara\'iso,
Av. Gran Breta\~na 1111, 5030 Casilla, Valpara\'iso, Chile}

\affiliation{N\'ucleo Milenio de Formaci\'on Planetaria - NPF,
Universidad de Valpara\'iso,
Av. Gran Breta\~na 1111, Valpara\'iso, Chile}

\author[0000-0003-2862-5363]{Carlos Carrasco-Gonz\'alez}
\affiliation{Instituto de Radioastronom\'ia y Astrof\'isica - IRyA, UNAM, campus Morelia, Antigua carretera a P\'atzcuaro N°. 8701. Col. Ex Hacienda San Jos\'e de la Huerta, Morelia, Michoac\'an, C.P. 58089, M\'exico}

\author[0000-0003-3903-8009]{Matthias R. Schreiber}
\affiliation{Universidad T\'ecnica Federico Santa Mar\'ia, UTFSM, Avda. Espa\~na 1680, Valpara\'iso, Chile}

\affiliation{N\'ucleo Milenio de Formaci\'on Planetaria - NPF,
Universidad de Valpara\'iso,
Av. Gran Breta\~na 1111, Valpara\'iso, Chile}

\author[0000-0002-5991-8073]{Anibal Sierra}
\affiliation{Universidad de Chile, Departamento de Astronom\'ia, Camino El Observatorio 1515, Las Condes, Santiago, Chile}

\author[0000-0003-4475-3605]{Johan Olofsson}
\affiliation{Instituto  de  F\'isica  y  Astronom\'ia,
Facultad  de  Ciencias,  Universidad de Valpara\'iso,
Av. Gran Breta\~na 1111, 5030 Casilla, Valpara\'iso, Chile}

\affiliation{N\'ucleo Milenio de Formaci\'on Planetaria - NPF,
Universidad de Valpara\'iso,
Av. Gran Breta\~na 1111, Valpara\'iso, Chile}

\author[0000-0001-7868-7031]{Amelia Bayo}
\affiliation{Instituto  de  F\'isica  y  Astronom\'ia,
Facultad  de  Ciencias,  Universidad de Valpara\'iso,
Av. Gran Breta\~na 1111, 5030 Casilla, Valpara\'iso, Chile}

\affiliation{N\'ucleo Milenio de Formaci\'on Planetaria - NPF,
Universidad de Valpara\'iso,
Av. Gran Breta\~na 1111, Valpara\'iso, Chile}

\author[0000-0002-6617-3823]{Claudio Caceres}
\affiliation{Departamento de Ciencias Fisicas, Facultad de Ciencias Exactas, Universidad Andres Bello, Av. Fernandez Concha 700, Las Condes, Santiago, Chile}

\affiliation{N\'ucleo Milenio de Formaci\'on Planetaria - NPF,
Universidad de Valpara\'iso,
Av. Gran Breta\~na 1111, Valpara\'iso, Chile}

\author[0000-0001-7668-8022]{Hector Canovas}
\affiliation{Telespazio UK for the European Space Agency (ESA), European Space Astronomy Centre (ESAC), Camino Bajo del Castillo s/n, 28692 Villanueva de la Ca\~nada, Madrid, Spain}

\author[0000-0002-9569-9234]{Aina Palau}
\affiliation{Instituto de Radioastronom\'ia y Astrof\'isica - IRyA, UNAM, campus Morelia, Antigua carretera a P\'atzcuaro N°. 8701. Col. Ex Hacienda San Jos\'e de la Huerta, Morelia, Michoac\'an, C.P. 58089, M\'exico}








\begin{abstract}

One of the most important questions in the field of planet formation is how mm-cm sized dust particles overcome the radial drift and fragmentation barriers to form kilometer-sized planetesimals. 
ALMA observations of protoplanetary disks, in particular transition disks or disks with clear signs of substructures, can provide new constraints on theories of grain growth and planetesimal formation and therefore represent one possibility to progress on this issue. 
We here present ALMA band 4 (2.1 mm) observations of the transition disk system Sz\,91 and combine them with previously obtained band 6 (1.3 mm) and 7 (0.9 mm) observations. Sz\,91 with its well defined mm-ring, more extended gas disk, and evidence of smaller dust particles close to the star, is a clear case of dust filtering and the accumulation of mm sized particles in a gas pressure bump. 
We computed the spectral index (nearly constant at $\sim$3.34), optical depth (marginally optically thick), and maximum grain size ($\sim\,0.61$\,mm) in the dust ring from the multi-wavelength ALMA observations and compared the results with recently published simulations of grain growth in disk substructures. Our observational results are in very good agreement with the predictions of models for grain growth in dust rings that include fragmentation and planetesimal formation through the streaming instability.

\end{abstract}

\keywords{protoplanetary disks --- stars: individual: Sz\,91 --- stars: variables: TTauri 
}


\section{Introduction} \label{sec:intro}

The most critical step in our understanding of the formation of terrestrial planets and giant planet cores 
is the assembly of kilometer-sized planetesimals from smaller dust particles \citep[e.g.][]{johansen14}. 
The key problem is that the predicted short radial drift time scales for mm/cm sized particles when they decouple from the sub-Keplerian gas flow strongly limits the possibility of these particles to grow into planetesimals \citep{weidenschilling77}.
The fact that planets exist and   
mm sized dust particles are routinely observed at distances of $\sim10-100$\,au  
from the host star \citep[e.g.][]{barenfeld17,ansdell2018,long2019,tazzari2020a,cieza2020} clearly shows that the predicted fast inward migration of these solids is suppressed in most disks. 

A crucial assumption leading to the predicted inward drift is that the gas pressure 
in the disk continuously decreases with radius. 
Recent high resolution ALMA observations 
of protoplanetary disks are indicating that this assumption is most likely not generally correct. 
The ground-breaking image of the HL\,Tau disk \citep{ALMA15} and the results of the DSHARP program \citep{andrews2018} revealed 
that ring-like dust sub-structures are ubiquitous in relatively young protoplanetary disks. These findings are 
complemented by the identification of transition disks that show dense dust rings around dust depleted cavities \citep[e.g.][]{andrews11a,pinilla2018,vandermarel2018}. In several cases these ring-like structures are accompanied by azimuthal asymmetries \citep[e.g.][]{casassus2013,vandermarel2021}, spiral arms \citep[e.g.][]{christiaens2014,huang2018b}, or small and possibly inclined inner disks deduced from NIR observations \citep[e.g.][]{marino2015} or using ALMA data \citep[e.g.][]{perez2018, francis2020}.

One explanation for the observed ring-like sub-structures is that dust accumulates at local gas pressure maximums. 
These substructures in the gas pressure distribution may solve the drift and planetesimal formation timescale problems.
For the origin of these pressure bumps in transition disks embedded planets or companions \citep[e.g.,][]{zhu2011}, dead-zones \citep[e.g.,][]{flock15}, and/or internal photoevaporation \citep{alexander_armitage07} have been suggested. Photoevaporation alone is generally ruled out by the high accretion rates found in most transition disks \citep{owen_clarke2012}, and dead zones alone fail to explain deep observed gas gaps \citep{pinilla2016a}.
Planet-disk interactions are particularly favored as they can explain some of the main features observed in most systems: differences in radii between the gas and the dust component and also between dust grains of different sizes \citep[e.g.,][]{garufi2013,vandermarel2016,dong17}, for instance. If Jupiter-like planets are responsible for the pressure bump, then  the small particles will leak into the cavity reaching inner regions \citep[e.g.,][]{pinilla2016a}. This allows for the presence of inner disks over million years timescales \citep[e.g.,][]{pinilla2021}.
In any case, the mm/cm sized particles that are trapped in the pressure bump may have enough time to agglomerate 
and form planetesimals. 

Grain growth significantly affects the optical properties of dust particles. At millimeter wavelengths the slope of the spectral energy distribution, i.e. the spectral index, has been widely used to estimate the maximum grain size of dust grains \citep[e.g.,][]{williams_cieza_2011}. Several observational studies have now provided strong evidence for grain growth in disks \citep[e.g.,][]{calvet02,testi03,natta04,perez12,tsukagoshi2016,carrasco2019}. These observational results have recently been complemented with simulations of grain growth in a dust ring \citep{stammler2019} including fragmentation and planetesimal formation. 
According to these models, given the self-regulating nature of planetesimal formation which stabilizes the midplane dust-to-gas ratio, once fragmentation sets in, the spectral index in the ring is predicted to reach its minimum value. Furthermore, in line with the idea that planetesimal formation may take place in ring-like substructures, the simulations show that planetesimal formation can naturally explain the surprisingly similar 
optical depths derived for the rings of the DSHARP targets \citep{dullemond2018,huang2018a}.    
However, given the still small number of targets observed at high resolution,  
it cannot be excluded that the derived optical depths cluster in the range of 0.2-0.6 by coincidence \citep{stammler2019}.   
Moreover, it remains to be tested if the ring-like accumulations around the dust
depleted cavities in transition disks show similar characteristics as the sub-structures of the DSHARP disks. 
Thus, additional high-resolution observations of ring structures are highly required. In this regard, Sz\,91 represents an ideal complementary target to the DSHARP sample. 

Sz\,91 is a 3-5\,Myr old T\,Tauri star of spectral type M0 located in the Lupus III molecular cloud \citep{romero2012,canovas2015a} at a distance of 159 $\pm$ 2 pc \citep[Gaia EDR3;][]{gaia_mission2016,gaia_EDR32020} accreting from a transition disk at a rate of 
$\dot{M} \sim\,10^{-8.8} \rm M_{\odot}\,yr^{-1}$ \citep{alcala2017}. 
High angular resolution ALMA band 7 observations revealed that the mm-sized dust particles are concentrated in a ring-like structure \citep{canovas2016} extending from 86 to 101 au \citep{francis2020} while polarized light observations are best explained by small, porous grains distributed in a disk with a significantly smaller ($\sim45$\,au) cavity \citep{mauco2020}. The gas component of the disk extends from inside the mm-cavity, with a gas-depleted cavity at 37 au \citep{vandermarel2021}, up to $\sim\,400$\,au from the star \citep{tsukagoshi2019}. 
Such a disk structure, an extended gas disk, a concentration of mm-sized particles in a 
well defined ring-like structure, and smaller dust particles inside the mm cavity is indicative of radial drift of the mm-sized dust particles that decoupled from the gas flow and the presence of a local pressure maximum that acts as a dust filter. This filter halts the radial drift of the mm/cm sized particles while the smaller dust particles pass the pressure maximum \citep[e.g.][]{rice2006}. 

Here we present new ALMA band 4 (2.1 mm) observations of the transition disk Sz\,91 and combine them with the archival band 6 (1.3 mm) and band 7 (0.9 mm) data to constrain the optical depth, the dust surface density and the grain size distribution of the dust particles that are trapped in the ring by performing a radial analysis of the (ALMA) spectral energy distribution (SED) including optical depth and scattering effects. Comparing our results with predictions of theoretical models of grain growth in ring-like substructures that include planetesimal formation \citep{stammler2019}, we find excellent agreement with our observational results which may indicate that planetesimal formation is occurring in the mm dust ring around Sz\,91. 

\section{Observation and data reduction} \label{sec:obs}

With the aim of characterizing the dust content in the ring around Sz\,91, we analysed new 2.1 mm ALMA data (band 4) as well as archival 1.3 mm (band 6) and 0.9 mm (band 7) data. 

\subsection{Band 4 observations}

We obtained 2.1 millimeter observations of Sz\,91 (project ID 2018.1.01020.S) by combining
ALMA 12-m array extended (C43-8) and more compact
(C43-5) configurations, resulting in baselines ranging from
15 meters to 8.55 kilometers and a total of 43-48 antennas. The combined observations are sensitive to spatial scales
of up to 17{\arcsec}. The long baseline observations were acquired in three different blocks of $\sim$50 min each (2.15h in total) on July 22, 23, and 28 in 2019 (cycle 6). The short baseline observations were executed in two different blocks of $\sim$40 min each ($\sim$1.3h in total) on November 11 and 22 in 2018.

Precipitable water vapor ranged between 0.5 and 2.0 mm.
Observations of a phase calibrator (J1610-3958) were alternated with the science observations to calibrate the time-dependent variation of the complex gains.
The cycling time for phase calibration was set to 8 minutes and 54 seconds for the compact and extended configurations, respectively.
The ALMA correlator was configured in Frequency Division Mode (FDM). The band 4 receiver system was employed to detect the continuum emission at band 4 (146.40 GHz) where four spectral windows with 1.875 GHz bandwidth were set up for detecting the dust continuum, centered at 132.4 GHz, 134.35 GHz, 144.46 GHz, and 146.40 GHz, respectively. We set up the frequency of the last spectral window aiming for a possible detection of the CS v=0 J=3-2 spectral line at 146.9690 GHz with a channel spacing of 976.562 kHz, which corresponds to a velocity resolution of $\sim 2.0$ km/s. 

The visibility data were reduced and calibrated using the
Common Astronomical Software Application package \citep[CASA;][]{McMullin2007}.
The raw data were calibrated with the reduction script provided by 
the ALMA staff -- Pipeline version 42030M (Pipeline-CASA54-P1-B), 
which includes offline Water Vapor Radiometer (WVR) calibration, system temperature correction, as well as bandpass, phase, and amplitude calibrations. The short baseline and long baseline data-sets were calibrated independently. We shifted the phase center of the visibilities of the compact configuration to the position of Sz\,91 at the date of the long baseline observations, as determined from the position and proper motions  of the source (-10.0, -22.8) mas yr$^{-1}$ reported by Gaia EDR3 \citep{gaia_mission2016,gaia_EDR32020}, using the task \texttt{fixvis} in CASA.

The image reconstruction was performed using the 
CLEAN algorithm (CASA version 5.6.2-2, task \texttt{tclean}) 
using Briggs weighting with a robust parameter of 0.5.
The final continuum image has a peak of 70 $\mu$Jy beam$^{-1}$ and an rms of 5 $\mu$Jy beam$^{-1}$ for a synthesized beam of 103$\times$91 mas. Integrating the flux inside an elliptical region enclosing the totality of the source with semi-major axis $a = 1.03 \arcsec$ and semi-minor axis $b = 0.93 \arcsec$ results in $F_{\rm 2.1\,mm}$ = 2.3 $\pm$ 0.1 mJy.
We also obtained a continuum image at 220 mas resolution (see sec~\ref{sec:archival_data}), peaking at 0.16 mJy beam$^{-1}$ and with an rms of 0.05 mJy beam$^{-1}$. The integrated flux inside the same elliptical region as before results in $F_{\rm 2.1\,mm}$ = 2.1 $\pm$ 0.1 mJy. The integrated flux errors include the absolute flux calibration error of 5\% which dominates the uncertainty.

The visibilities of the line emission, on the other hand, were obtained by
subtracting the continuum visibilities with velocity steps of 2.0 km s$^{-1}$.
The CLEAN image was obtained using a natural weighting and has a beam size of 197\,$\times$\,180 mas with an rms noise
level of 0.32 mJy beam$^{-1}$. The CS v=0 J=3--2 line detection is discussed in the appendix.  

\subsection{Archival data} \label{sec:archival_data}

We used archive ALMA continuum images of Sz\,91 at band 7 (project ID 2012.1.00761.S) and band 6 (project ID 2015.1.01301.S) to complement our study in order to estimate spectral indices. For this, we obtained CLEAN images following the same methodology and parameters as for our band 4 observations. Table~\ref{tab:ALMA_img} lists the characteristics of the final ALMA images used in this work. 

To compare images at different wavelengths, they must be at the same angular resolution.
Given that the weighting of the visibilities affects the sensitivity and
angular resolution of the final images (more weighting to longer baselines implies a worse sensitivity), we compared the
beam sizes and rms noises obtained for different weightings at different bands and concluded that the best compromise
between resolution and sensitivity is obtained by using a Briggs weighting with a robust parameter of 0.5 and convolving all
images to a circular beam with a diameter of 220 mas, 
equivalent to a physical size of $\sim$35 au at the distance of Sz\,91 ($d$ = 159 $\pm$ 2 pc). 
Then, we obtained the corresponding cleaned images at each
wavelength with angular resolutions of 220 mas or less (see Table~\ref{tab:ALMA_img}) and
convolved them using the task \texttt{imsmooth} from CASA.

The final band 7 image peaks at 4.0 mJy beam$^{-1}$ and has an rms of 1.0 mJy beam$^{-1}$, and the band 6 image reaches 0.76 mJy beam$^{-1}$ and has an rms of 0.30 mJy beam$^{-1}$, both for a circular synthesized beam of 220 mas. We remark, however, that the whole continuum windows for the band 6 data-set was flagged due to problems with the correlator. Therefore, we only used the line-free channels of the spectral windows assigned for line detection to make the image. This is the reason for its low peak intensity. The synthesized beam of 220 mas is basically set by this (noisy) data-set. Integrating the flux inside the same region used for the B4 images results in $F_{\rm 0.9\,mm}$ = 44 $\pm$ 4 mJy and $F_{\rm 1.3\,mm}$ = 13.0 $\pm$ 0.6 mJy. These values include the absolute flux calibration error at each frequency. For this, we used the nominal values of 10\% at band 7, and 5\% at band 6. We remark that the integrated flux errors obtained here are dominated by the flux calibration errors.

\section{Results} \label{sec:results}

\subsection{Continuum Emission and Radial Profiles}

\begin{figure}
	\includegraphics[width=\columnwidth]{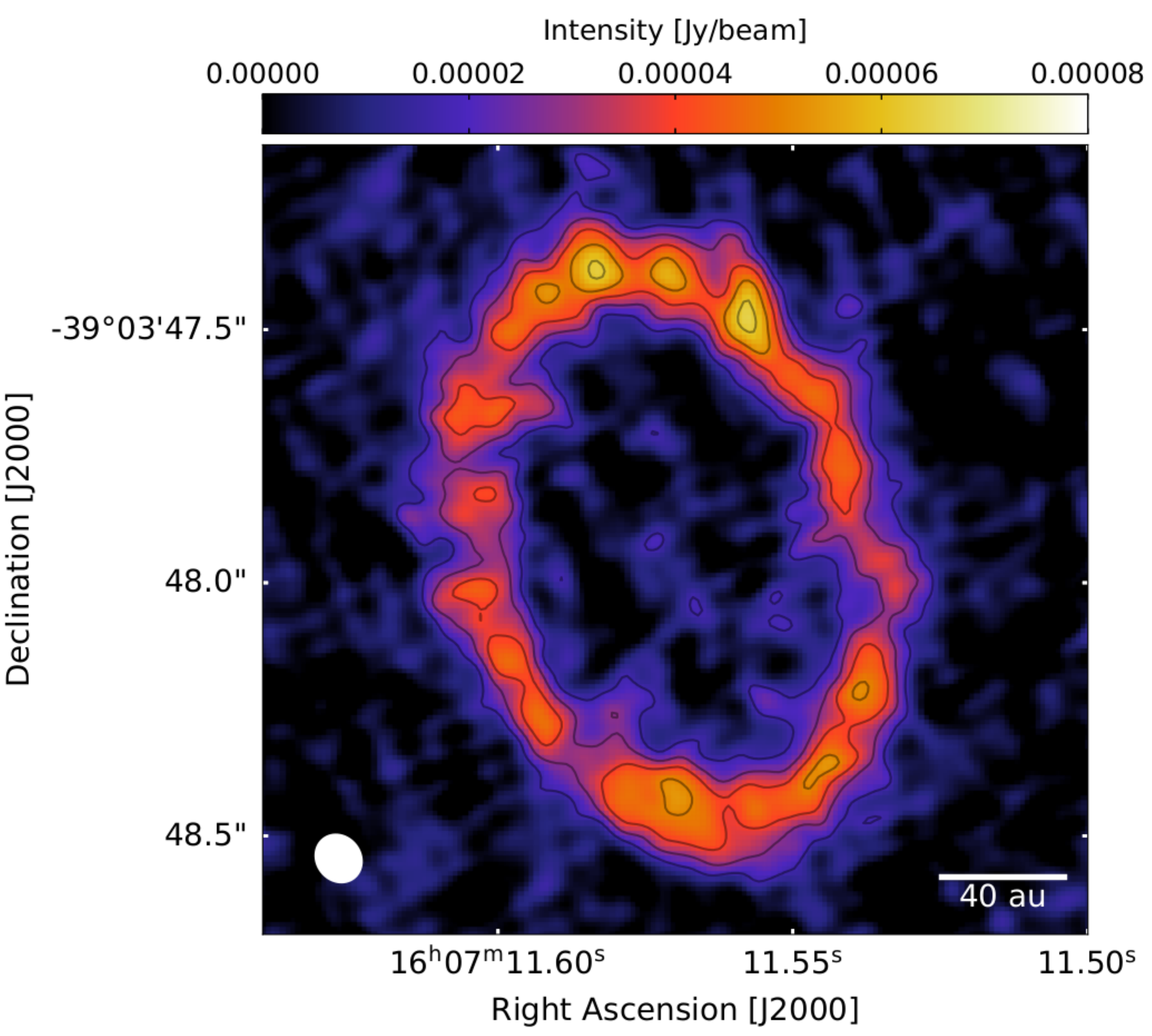}
    \caption{ALMA continuum observations at 2.1 mm (band 4) of the disk around Sz\,91. The beam size is 103\,$\times$\,91 mas (16.4\,$\times$\,14.5 au) and is shown in the bottom left of the image. Contour levels correspond to 3, 5, 7, 9, 11 $\sigma$, where $\sigma$ is the image RMS.} 
    \label{fig:B4_cont}
\end{figure}

Figure~\ref{fig:B4_cont} shows the 2.1 mm (band 4) ALMA continuum image of Sz\,91. We clearly resolved the disk ring structure, with the north side being brighter than the south side (peak-to-peak ratio of 1.15), something also found in ALMA observations at 0.9 mm in \citet{tsukagoshi2019}. 
Unfortunately, given the current resolution of our data we could only partially resolve the ring in the radial direction.

\begin{figure}
	\includegraphics[width=\columnwidth]{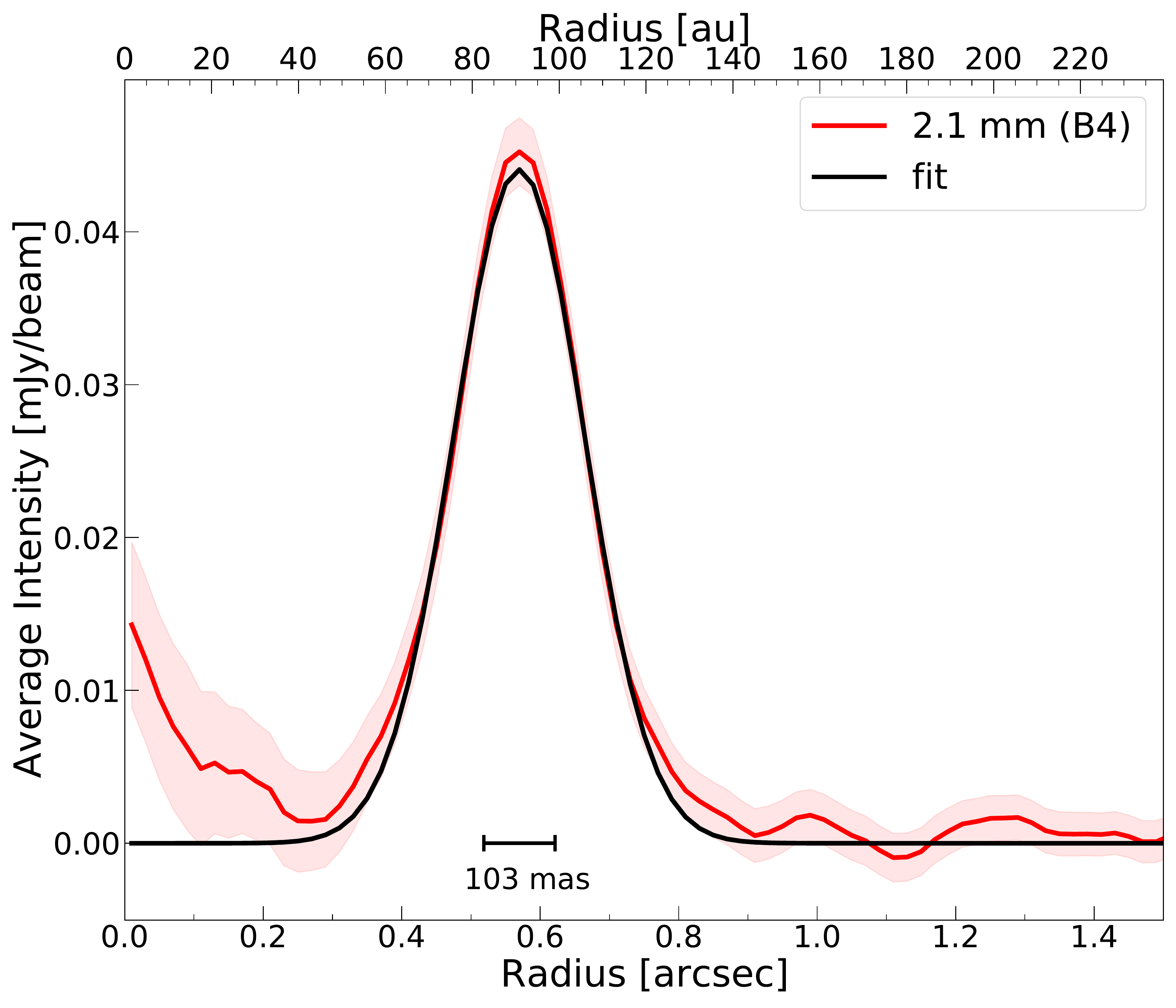}
    \caption{Radial profile (red line) after azimuthally averaging the deprojected continuum image shown on Figure~\ref{fig:B4_cont} (assuming d=159 pc) at 2.1 mm. Statistical errors (based on the rms noise of the map) at 1$\sigma$ level are shown as a filled band around the nominal value. A Gaussian fit to the profile is shown as a solid black line with a FWHM of 0.21\arcsec (33.4 au). The horizontal bar indicates the beam major axis for comparison.}
    \label{fig:Radprofiles}
\end{figure}

\begin{figure}
	\includegraphics[width=\columnwidth]{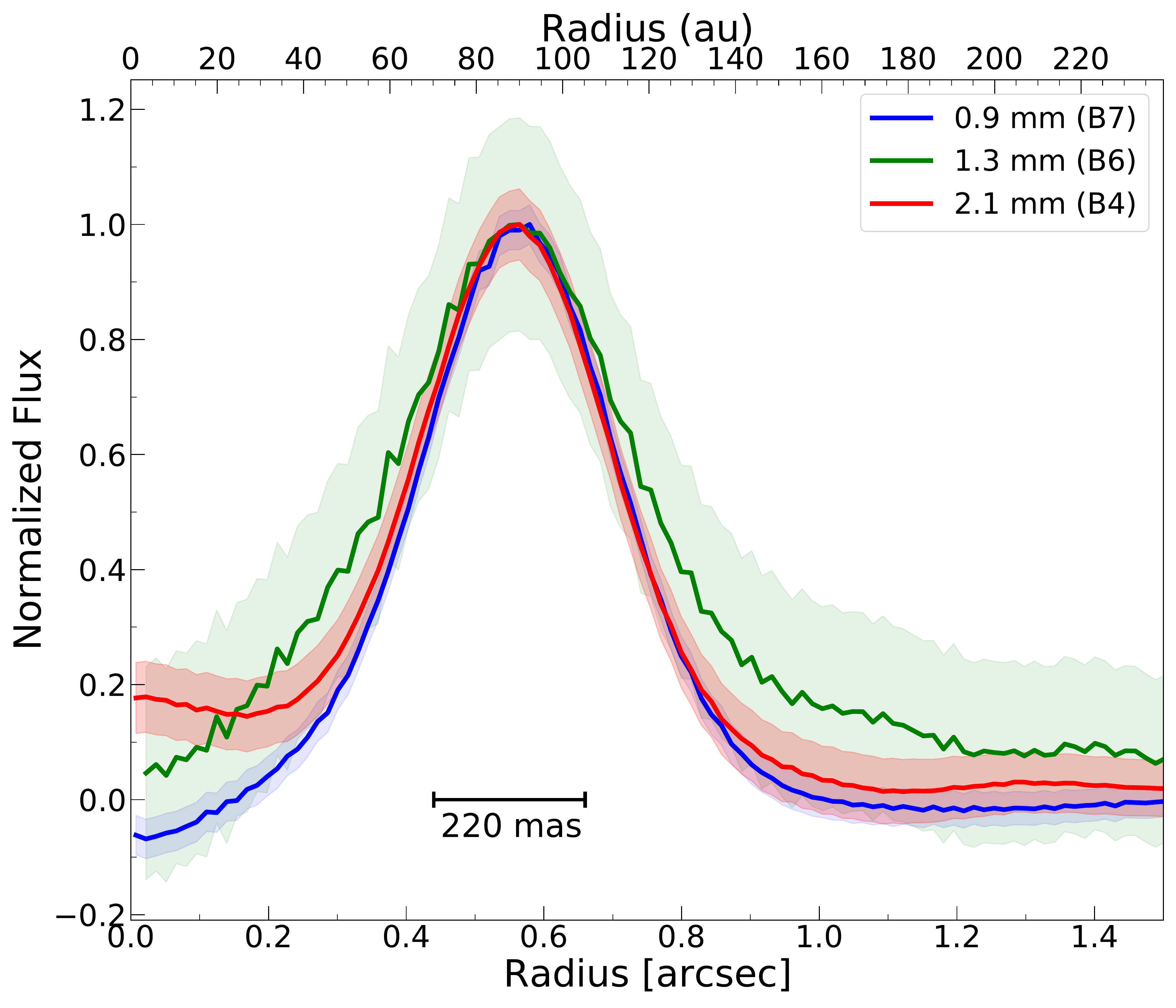}
    \caption{Normalized radial profiles of the three data-sets used in this work (see Table~\ref{tab:ALMA_img}) convolved to a circular beam of 220 mas (horizontal bar). Statistical errors (based on the rms noise of each map) at 1$\sigma$ level are shown as filled bands around the nominal values in each case.}
    \label{fig:Radprofiles_stack}
\end{figure}

We estimated the radial profiles of the dust emission at
different wavelengths by averaging emission in concentric elliptical rings with widths of 15 mas and with an eccentricity given by 
the disk inclination (49.7\degr) and position angle (18.1\degr) taken from the ALMA band 7 continuum image from \citet{tsukagoshi2019}. The intensity at each radius is given by the azimuthally-averaged
intensity in the ring, and the errors are calculated as
$\Delta I_{\nu}=\rm rms_{\nu}/(\Omega_{ring}/\Omega_{beam})^{0.5}$, where $\rm \Omega_{ring}$ and $\rm \Omega_{beam}$ are the
solid angles of the ring and the synthesized beam, respectively.
The brightness temperature profiles are calculated from the blackbody Planck function without assuming the Rayleigh-Jeans regime to avoid errors at short wavelengths in the outer parts of the disk where dust temperatures are expected to be low. The profiles are calculated through the following relationship:

\begin{equation}
    T_{B} = \frac{h \nu} {k_{B}\, \ln{\bigg[ 1+\frac{2h\nu^{3}} {I_{\nu}c^{2}} } \bigg ]  },
\end{equation}

\noindent
where $h$ is the Planck constant, $k_{B}$ is the Boltzmann constant, $I_{\nu}$ is the intensity at frequency $\nu$, and $c$ is the speed of light.  

The radial profile of our band 4 observations is shown in Figure~\ref{fig:Radprofiles} (red line). Statistical errors (based on the rms noise of the map) at 1$\sigma$ level are shown as shaded regions around the nominal value. We fitted a Gaussian curve to the profile using the python module \texttt{mpfit}\footnote{https://github.com/segasai/astrolibpy/blob/master/mpfit/mpfit.py} (black line) with a FWHM of 0.21\arcsec (33.4 au). 
On the other hand, the normalized radial profiles of all the three data-sets used in this work are shown in Figure~\ref{fig:Radprofiles_stack} for a circular synthesized beam of 220 mas (35 au).
The peak of emission is located at $\sim$90 AU from the central star in all the profiles. 

We remark that even though we use a data-set convolved to a resolution of 220 mas (where the disk is not radially resolved), this resolution is very similar to the actual width of the ring estimated with our highest angular resolution observations at band 4, as shown in Figure~\ref{fig:Radprofiles}. In this sense, the estimate of physical parameters reported in sec~\ref{sec:results},
despite being considered as lower limits, should be very similar to the average values of the real disk parameters.

The integrated flux we estimated at band 6 and band 7 for a resolution of 220 mas (sec~\ref{sec:archival_data}) are consistent with the reported flux in the literature. \citet{tsukagoshi2019} reported a $F_{\rm 0.9\,mm}$ = 45.2 $\pm$ 0.5 mJy, estimated inside the regions with S/N $>$ 3$\sigma$. Note that, the flux reported in \citet{tsukagoshi2019} is at a resolution of $\sim$130 mas which agrees very well with the flux estimated from our band 7 image at 150 mas of $F_{\rm 0.9\,mm}$ = 44.6 $\pm$ 4.4 mJy (see Table~\ref{tab:ALMA_img}). On the other hand, our integrated flux at band 6 agrees very well with the one reported in \citet{canovas2015a} with a value of 12.7 $\pm$ 1.9 mJy.

Note the excess of emission inward 40 au at 2.1 mm on Figure~\ref{fig:Radprofiles_stack}, also visible at a higher resolution in Figure~\ref{fig:Radprofiles}. 
Moreover, the band 4 emission from inside the ring of dust (r $<$ 60 au) is clearly exceeding the band 7 profile. 
Given that the excess is not present in the radial profiles of \citet{tsukagoshi2019} at band 7, where the disk is expected to be brighter specially at a lower resolution, we can conclude that the excess observed in our data represents a real detection.

Based on this detection and the non-detection in band 7 we can estimate an upper limit to the spectral index at the inner regions. We found a $\alpha_{2.1 - 0.9\,\rm mm}$ upper limit $\lesssim$ 2.8. Even though this limit does not rule out the presence of an inner disk (dust thermal emission) the most probable explanation for it is free-free emission coming from  ionized gas close to the central star. This could produce an increase in the 2.1 mm emission at the inner regions of the disk, which would result in an apparent excess. We are inclined to the free-free emission scenario because the band 7 observations are more favorable at detecting an inner disk as stated above. Also, there is no NIR excess emission in the SED of the source \citep{canovas2015a}. 
Given all these reasons, we can conclude that the excess observed in our data is more consistent with free-free emission. This type of emission has also been detected at similar wavelengths in the TW Hydra disk recently \citep{macias2021} and has been reported at 7 mm, 15 mm and centimeter wavelengths in other disks as well \citep[e.g.,][]{ubach2012,ubach2017,macias2018}. 
Radio flux monitoring at multiple epochs and at different wavelengths are needed in order to disentangle the physical mechanisms responsible for the excess of emission at mm wavelengths in Sz\,91 \citep{ubach2017}.

We fitted Gaussian curves to the intensity profiles using \texttt{mpfit} as before and found FWHM of 0.353\arcsec $\pm$ 0.007 and 0.329\arcsec $\pm$ 0.004 for band 4 and band 7 profiles, respectively. The slight difference is due to the excess of emission mentioned before in the band 4 profile.  
Observations at higher angular resolution and at longer wavelengths are needed in order to investigate whether there is any variation in the width of the dust ring at different wavelengths, as expected from dust evolutionary models \citep{pinilla2015b,powell2019} but yet undetected observationally \citep[e.g.,][]{norfolk2021}.

\subsection{Spectral Indices and Brightness Temperature}

\label{sec:spectral_indices}
\begin{table*}
	\begin{center}
	\caption{ALMA images used to estimate spectral indices}
	\label{tab:ALMA_img}
	\begin{tabular}{lcccccccl} 
		\hline
		\hline
		 & Wavelength & Frequency & Maximum Intensity & Flux & SNR & RMS Noise & Resolution & Method \\
        Band & (mm) & (GHz) & (mJy beam$^{-1}$) & (mJy) & & (mJy beam$^{-1}$) & (mas) & \\
        \hline
        B7 & 0.9 & 349.4 & 4.0 & 44.0 & $\sim$33 & 1.0 & 220 & Radial profiles \\
        B6 & 1.3 & 225.0 & 0.76 & 13.0 & $\sim$7 & 0.30 & 220 & Radial profiles \\
        B4 & 2.1 & 139.4 & 0.16 & 2.1 & $\sim$20 & 0.05 & 220 & Radial profiles \\
		\hline
		B7 & 0.9 & 349.4 & 2.2 & 44.6 & $\sim$23 & 0.6 & 150 & CASA task \texttt{immath}\\
		B4 & 2.1 & 139.4 & 0.10 & 2.1 & $\sim$17 & 0.02 & 150 & CASA task \texttt{immath}\\
		\hline
	\end{tabular}
	\end{center}
    \begin{footnotesize}
{\bf Notes.} Band 6 continuum windows were completely flagged in the raw data-set due to problems with the correlator. \\
\end{footnotesize}
\end{table*}

The millimeter spectral index between two different wavelengths, defined as $\alpha = \log(I_{\nu1}/I_{\nu2})/\log(\nu_1/\nu_2)$, has been widely used to study grain growth in protoplanetary disks \citep[e.g.,][]{williams_cieza_2011}. 

\begin{figure*}
	\includegraphics[width=\textwidth]{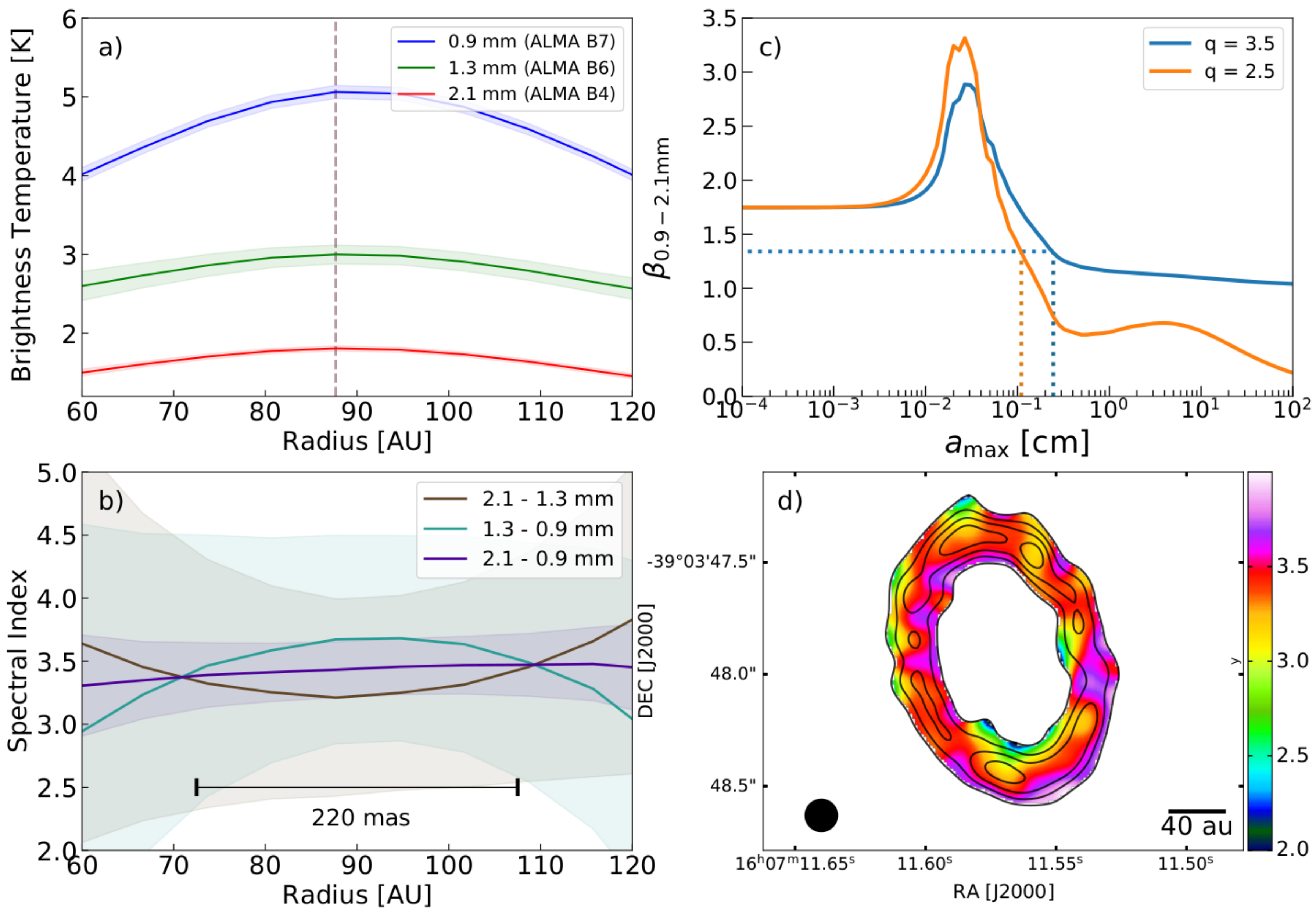}
	
    \caption{\textit{a)}: Radial profiles of the brightness temperature in the ALMA images. Dashed lines indicate brightness temperature peak values (same for all bands). 
    \textit{b)}: Spectral indices between different combinations of wavelengths. The spectral indices were computed using the radial profiles of Figure~\ref{fig:Radprofiles_stack} (at 220 mas resolution). Spectral indices are reliable only from 60 to 120 au where emission from the dusty disk is expected. Uncertainties at 1 $\sigma$ level are shown as shaded regions. We obtain an average spectral index of $\alpha$=3.34 $\pm$ 0.26.
    \textit{c)}: Slope of the dust opacity coefficient, $\beta$, from 0.9 - 2.1 mm using the DSHARP opacities from \citet{birnstiel2018} for a power-law index of the dust size distribution of q = 3.5 (blue) and q = 2.5 (orange). The horizontal dash line indicates the value of $\beta$ = 1.34 estimated for the disk around Sz\,91 (assuming optically thin emission). We obtained $a_{\rm max}$ values of $\sim$2.5 mm and $\sim$1 mm for q = 3.5, and q = 2.5, respectively (vertical dash lines).
    \textit{d)}: Spectral index map (at 5$\sigma$) between 2.1 and 0.9 mm using the final continuum images listed at the bottom of Table~\ref{tab:ALMA_img} convolved to an angular resolution of 150 mas (23.8 au) and shown by the filled circle at the bottom left. Overlaid are the contours of the continuum image at 2.1 mm (band 4) from Figure~\ref{fig:B4_cont} at the same resolution. Contour levels correspond to 5, 8, 11, and 14 $\sigma$, where $\sigma$ is the image RMS. The spectral index has an average value of 3.37.}
    \label{fig:Tb_alpha}
\end{figure*}

We derived spectral indices based on the band 4, 6, and, 7 ALMA data. Using the intensities observed at different wavelengths, taken from their respective radial profiles (at 220 mas resolution), we obtained spectral indices by combining data at multiple bands: $\alpha$(2.1-1.3 mm), $\alpha$(1.3-0.9 mm), and $\alpha$(2.1-0.9 mm). 
Figure~\ref{fig:Tb_alpha} shows the brightness temperatures (panel $a$) and spectral indices (panel $b$) estimated this way. As shown in the figure, the spectral index throughout the disk region is almost constant with an average value of 3.34 $\pm$ 0.26. This result is consistent with the value of $\alpha_{\rm 0.8-2.7mm}$ = 3.36 found previously by \citet{canovas2015a}. The $x$-axis in panel $b$ is shown from 60 - 120 au since it is where the emission from the dusty ring is expected (see Figure~\ref{fig:Radprofiles}). The $\alpha$(2.1-0.9 mm) index is the most reliable index given its lower uncertainty level (shaded region). 

Based on this, and since the band 6 data-set has the lowest SNR,
we computed the spectral index map of $\alpha$(2.1-0.9 mm) by using the band 4 and 7 observations at a higher resolution. We used the CASA task \texttt{immath} with the mode option \textit{spix}. First, we constructed primary beam corrected images with a resolution of 150 mas (circular beam) using \texttt{imsmooth} as before (bottom part of Table~\ref{tab:ALMA_img}). This resolution is set by the band 7 image. Then, we aligned both data-sets using the CASA task \texttt{UVfix} where the band 7 visibilities were shifted in order to match those of the band 4 data-set. For this, we used the source proper motions which yielded offsets of $\Delta\alpha =0.04{\arcsec}$ and $\Delta\delta = 0.09{\arcsec}$. 

The result is an spectral index map as shown in Figure~\ref{fig:Tb_alpha} (panel $d$), in which we applied a filter to keep only the pixels with emission well above the noise level (5$\sigma$, 30 $\mu$Jy beam$^{-1}$) of the band 4 continuum image (see Table~\ref{tab:ALMA_img}). Neither a clear trend of increasing spectral index with radius was found (as expected from grain growth and radial drift) nor any azimuthal variations. However, we obtain a mean $\alpha$ value of 3.37, consistent with the $\alpha$ estimated from the radial intensity profiles. Furthermore, the flat shape of the spectral index observed in panel $b$ is also found when using the radial profiles of band 4 and band 7 at 150 mas. Higher resolution observations are needed in order to resolve the disk in the radial direction to be able to study radial modulations of $\alpha$.

Finally, the low brightness temperatures in Figure~\ref{fig:Tb_alpha} seem to indicate that the emission is optically thin. In this regime, the brightness temperature is given by $T_{B} = T_{d}$ $\tau_{\nu}$, where $T_{d}$ is the dust temperature and $\tau_{\nu}$ is the optical depth at frequency $\nu$. Since $\tau_{\nu}$ is $<$1 (i.e., optically thin), the $T_{B}$ that we detect is lower than the real dust temperature at that location.
The dust emission comes from regions located beyond 60 au from the central star (see Figures~\ref{fig:Radprofiles} and \ref{fig:Radprofiles_stack}), so low temperatures are expected at those large radii. Low brightness temperatures of roughly the same order (at similar radii) as the ones we found here have also been found in the past for similar objects \citep[e.g.][]{perez2015}. 

\subsection{Maximum grain size and optical depth}
\label{sec: amax and tau}

Dust grains are the main source of opacity in disks. Consequently, the emission and absorption of radiation by dust grains produces the final disk spectrum. 
The spectral index has been widely used to estimate grain properties, particularly the maximum grain size, $a_{\rm max}$, 
assuming that the emission from the disk is optically thin and that scattering can be neglected \citep[e.g.][]{Beckwith_Sargen_1991,jorgensen_2007}. 
However, already in an early work, \citet[][]{miyake_nakagawa1993} pointed out that the scattering coefficient is much larger than the absorption coefficient at millimeter wavelengths if the dust grains are mm-cm in size, which is expected to be the case in protoplanetary disks.
The importance of scattering at mm-cm wavelengths has been recently reinforced by other authors \citep{sierra2020,sierra2019,carrasco2019,liu2019,soon2017}.
The general picture that is emerging is that the assumption of optically thin emission with negligible contributions from scattering are not generally justified 
and that in order to use the spectral index to characterize dust grains, parameters such as the dust opacity and the scattering efficiency
need to be taken into account. 

In what follows we estimate the maximum grain size in the ring around Sz\,91 in two different ways. First, we follow the classical approach and assume that the disk is optically thin and that scattering can be neglected. We complement this simple estimate by performing a radial analysis of the mm spectrum to simultaneously obtain the dust surface density, the optical depth, and the maximum particle size at each radius without assuming any value of the optical depth at any wavelength, and including scattering effects. 

\subsubsection{Classical Approach} \label{sec:classical amax}

At millimeter wavelengths, the emission is within the Rayleigh-Jeans regime so the emergent intensity is proportional to the Planck function, $B_{\nu}(T_d)$, which in turn behaves as $B_{\nu}(T_d) \propto \nu ^{2}$. Also, at these longer wavelengths, the opacity follows a power-law dependency on frequency. In the optically thin regime ($\tau_{\nu}$ $\ll$ 1) and in a pure absorption case, the total intensity can be written as $I_{\nu}\propto \nu^{2+\beta_{\rm abs}}$, where $\beta_{\rm abs}$ is the slope of the absorption coefficient. 
Therefore, knowing the spectral slope $\alpha$ = 2+$\beta_{\rm abs}$, i.e., the spectral index, allows to infer the maximum grain size, $a_{\rm max}$, from the spectral behavior of $\beta_{\rm abs}$ for different particle size distributions. 
The latter is typically assumed 
to be adequately described by a power-law of the form $n(a) \propto a^{-q}$, with $a$ as the particle radius, and with slope $q = 3.5$, resembling the size distribution found in the ISM. 
However, lower values of $q$ are expected in protoplanetary disks as grain growth acts in the system building up mm-cm particles \citep{drazkowska2019}. 
We therefore used, in addition to the canonical value ($q=3.5$), a smaller exponent of $q=2.5$. 
For these assumptions and using the DSHARP opacities \citep{birnstiel2018}, our spectral index value of $\alpha$ = 3.34 found in section~\ref{sec:spectral_indices} will translate to $\beta_{\rm abs}$ = 1.34 which implies $a_{\rm max}$ $\sim$ 2.5 mm and 1 mm for $q=3.5$ and $q=2.5$, respectively (panel $c$ in Figure~\ref{fig:Tb_alpha}).  

\subsubsection{Radial Fitting} \label{sec:radial_SED_fit}

Given that protoplanetary disks can be optically thick and 
that scattering can dominate the total emission, we complement the simple estimate of the maximum particle 
size by using the approach described in detail in \citet{carrasco2019}. 
We perform a radial analysis of the mm spectrum by modeling the radial intensity profiles at each ALMA wavelength assuming an axisymmetric, geometrically thin, and vertically isothermal disk. We also assume a constant dust temperature along the line of sight.
Since dust evolution models predict slightly flatter slopes for the particle size distribution at the position of dust rings \citep[e.g.,][]{drazkowska2019}, we assume $q=3.0$ and adopt the dust composition used by the DSHARP program \citep{birnstiel2018}.
The mm spectrum at each radius is fitted using the dust continuum emission at 0.9, 1.3, and 2.1 mm, up to a radius where the observations at all bands have a SNR of at least 1.5.
In our modeling, the dust scattering effects on the radiative transfer equation \citep[][]{zhu2019} are taken into account. In particular, we use the solution found by \citet{sierra2019}, which was also used to fit the dust properties in the disk around HL Tau \citep{carrasco2019}. According to \citet{sierra2019}, the emergent intensity at a particular radius can be written as:

\begin{equation}
    I_{\nu} = B_{\nu}(T_d)[(1-\exp(-\tau_{\nu}/\mu)) + \omega_{\nu}F(\tau_{\nu},\omega_{\nu})],
\end{equation}

\noindent
where $\tau_{\nu} = \Sigma_{d} \chi_{\nu}$ is the optical depth, $\Sigma_{d}$ is the dust surface density, $\chi_{\nu} = \kappa_{\nu} + \sigma_{\nu}$ is the total dust opacity (i.e., the extinction coefficient) with $\kappa_{\nu}$ and $\sigma_{\nu}$ as the absorption and scattering coefficient, respectively,   $\omega_{\nu} =\sigma_{\nu}/(\kappa_{\nu}+\sigma_{\nu})$ is the dust albedo, $\mu = \cos(i)$ is the cosine of the inclination angle, $i$, and $F(\tau_{\nu},\omega_{\nu})$ is defined as:

\begin{equation}
    \begin{aligned}
    F(\tau_{\nu},\omega_{\nu}) = \frac{1}{\exp(-\sqrt{3}\epsilon_{\nu}\tau_{\nu})(\epsilon_{\nu}-1)- (\epsilon_{\nu}+1)} \\  
    \times 
    \bigg [ \frac{1-\exp(-(\sqrt{3}\epsilon_{\nu}+1/\mu)\tau_{\nu})}{\sqrt{3}\epsilon_{\nu}\mu +1} \\
    + \frac{\exp(-\tau_{\nu}/\mu)-\exp(-\sqrt{3}\epsilon_{\nu}\tau_{\nu})}{\sqrt{3}\epsilon_{\nu}\mu - 1} \bigg ],
    \end{aligned}
\end{equation}

\noindent
where $\epsilon_{\nu} = \sqrt{1 - \omega_{\nu}}$. We are also including anisotropic scattering by using instead of $\sigma_{\nu}$ an effective scattering coefficient defined as $\sigma_{\nu}^{\rm eff} = (1-g_{\nu})\sigma_{\nu}$, where $g_{\nu}$ is the asymmetry parameter. 
Given a particle size distribution (with $q = 3.0$) and grain composition (from the DSHARP program), equation (2) depends only on three free parameters: $T_{d},\,\Sigma_{d},\,a_{\rm max}$. However, we fixed the dust temperature at each radius to the expected value for a passively irradiated flared disk in radiative equilibrium following equation (3) in \cite{huang2018a}. Therefore, our free parameters are just the dust surface density $\Sigma_{d}$, and the maximum grain size $a_{\rm max}$. For our analysis we vary $\Sigma_{d}$ from 10$^{-6.0}$ to 10 g/cm$^{2}$ and $a_{\rm max}$ from $10^{-1.7}$ to 1 cm, both in logarithmic space with a total of 100 intervals.

The probability of each parameter (at each radius) is computed by comparing the observed intensities with the emergent intensities for different combination of the free parameters. The probability $P$ for each pair of the free parameters is computed using a log-normal likelihood function:

\begin{equation}
    P = \exp \left ( -0.5 \, \sum_{i} \left ( \frac{I_{i} - I_{m,i}}{\hat{\sigma}_{I,i}} \right)^{2} \right ),
\end{equation}

\noindent
where $I_{i}$ is the azimuthally averaged intensity for a given radius and frequency, $I_{m,i}$ is the model intensity at the same radius and frequency, and $\hat{\sigma}_{I,i}$ is the uncertainty (also at the same radius) given by:

\begin{equation}
    \hat{\sigma}_{I,i} = \sqrt{\sigma_{I,i}^{2} + (\delta_{i}I_{i})^2},
\end{equation}

\noindent
where $\sigma_{I,i}$ is the error of the mean computed from the azimuthally averaged intensity profiles, and $\delta_{i}$ is the absolute flux calibration error at each frequency. For this, we used the nominal values of 10\% at band 7, and 5\% at bands 6 and 4. 
A detailed description and discussion of this methodology is presented in \citet{sierra2021}.
The radial profiles of the dust surface density and maximum grain size are constructed using the free parameters with the highest probability at each radius.

\begin{figure}
	\includegraphics[width=\columnwidth]{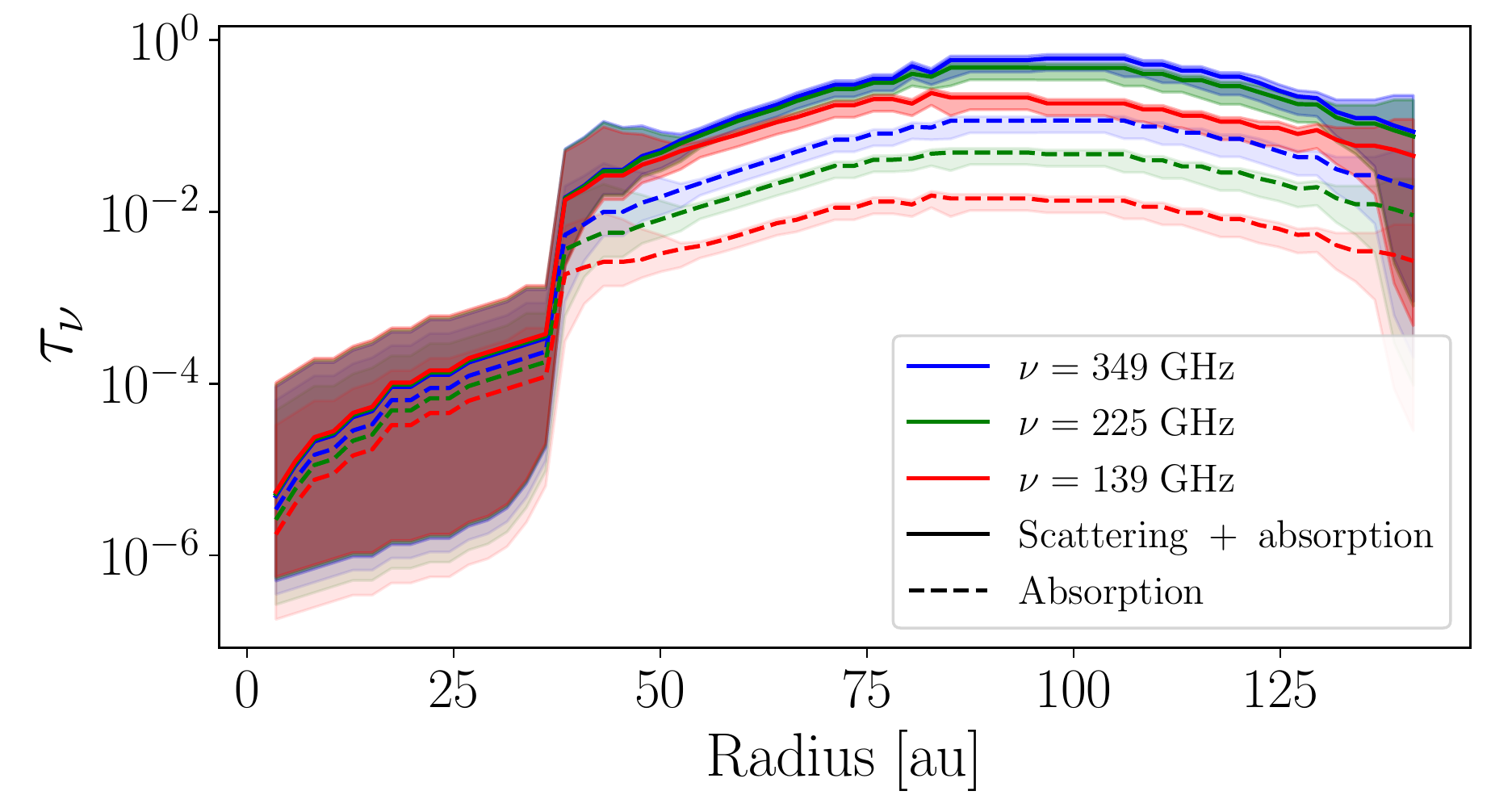}
    \caption{Optical depth profiles for the ALMA band 4 (red), band 6 (green), and band 7 (blue) observations. Dashed lines indicate pure absorption cases while solid lines show optical depths for cases also including scattering. 2 $\sigma$ level uncertainties are shown as shaded regions.}
    \label{fig:tau_dist}
\end{figure}

\begin{figure*}
	\includegraphics[width=\textwidth]{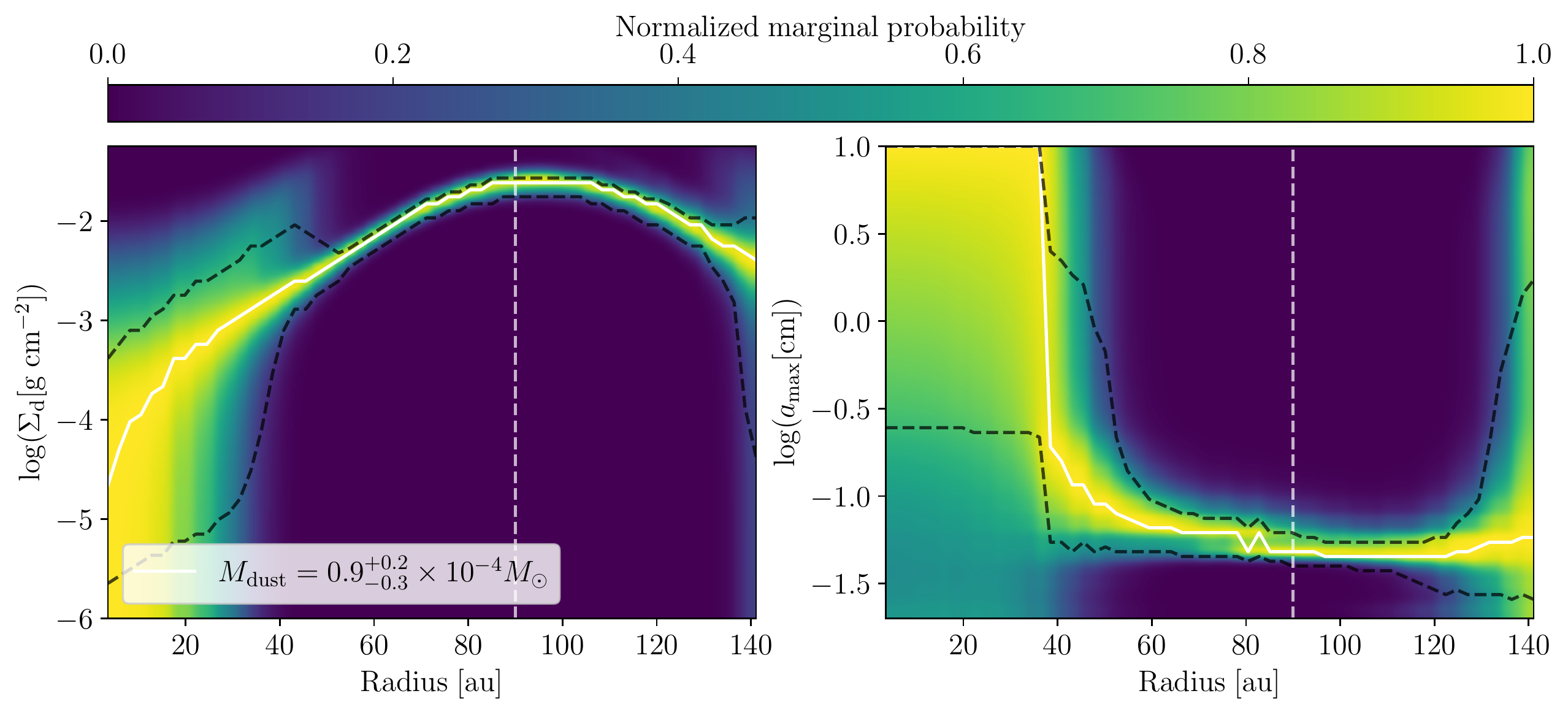}
    \caption{Radial fitting of the mm spectrum (see Section~\ref{sec: amax and tau} for details). Probability distributions of the surface density (left) and maximum grain size (right) are shown as a function of radius, where the colorbar shows the marginal probability. White solid lines indicate the best fit. Black dashed lines represent 2 $\sigma$ level uncertainties. Vertical lines indicate the location where the dust emission peaks (90 au). The total dust mass in the ring is 31.3 $M_{\oplus}$ and the maximum grain size $\sim$ 0.61 mm.}
    \label{fig:amax_dist}
\end{figure*}

The optical depth profiles obtained for the ring around Sz\,91 are shown in Figure~\ref{fig:tau_dist}. 
Red, green and blue lines indicate the ALMA band 4, 6, and 7 optical depths, respectively. Dashed lines represent the pure absorption case while solid lines trace cases including scattering. 2-$\sigma$ level uncertainties are shown as shaded regions. The emission is at least marginally optically thick for the more general case (with scattering) with peak values of $\tau_{\nu}$ decreasing to $\sim$0.01 for the longest wavelength if scattering is excluded. 
The resulting grain size and dust surface density profiles, on the other hand, are shown in Figure~\ref{fig:amax_dist}. White solid line highlights the most probable profile in both cases. The dust ring seems to be composed of particles with a maximum grain size of $\sim$ 0.61 mm. We remark that, similar to the case of the spectral index of Figure~\ref{fig:Tb_alpha}, the $a_{\rm max}$ estimate is only reliable from 60 to 120 au where the emission from the dusty disk is expected. 

A similar analysis but using a slope for the particle size distribution of $q = 2.5$ and $q = 3.5$ is shown in the appendix. Overall, we did not find a significant change in $a_{\rm max}$, which has an average value of 0.56 mm and 0.65 mm for $q = 2.5$ and $q = 3.5$, respectively. 
It is worth noting that, by ignoring scattering effects and assuming the disk to be optically thin could easily lead to overestimation of the dust particle sizes in protoplanetary disks as shown by our more realistic estimates of $a_{\rm max}$, which are from $\sim$2 and up to 4 times smaller than the $a_{\rm max}$ obtained through the classical approach in section~\ref{sec:classical amax}.

\subsection{Dust mass}

Using the dust surface density obtained with our multi-wavelength analysis described in section~\ref{sec:radial_SED_fit}, we can estimate the total dust mass in the disk around Sz\,91 by integrating the $\Sigma_{d}$ profile of Figure~\ref{fig:amax_dist} over the disk area.
By doing so, we found a total dust mass of $M_{\rm dust}$ = 31.3 $^{+6.3}_{-10.6}$ $M_{\oplus}$ (also shown in the left panel of Figure~\ref{fig:amax_dist} in Solar masses). Our dust mass estimate is 3.4 times larger than the value found in \citet{canovas2015a} of $M_{\rm dust}$ = 9.08 $M_{\oplus}$ (scaled to the same distance), estimated from fitting the entire SED (from the optical to the millimeter) assuming an $a_{\rm max}$ = 1 mm. On the other hand, our $M_{\rm dust}$ is $\sim$2 times higher than the dust mass estimated in \citet{vandermarel2018} with a value of $M_{\rm dust}$ = 15.8 $M_{\oplus}$ (also scaled to the true distance).

In order to compare to dust masses obtained from sub-mm surveys, where the dust emission is usually assumed to be optically thin, we also estimate $M_{\rm dust}$ using the simplified relationship of \citet{hildebrand83} that correlates the dust mass and the millimeter continuum flux as:

\begin{equation}
    M_{\rm dust} = \frac{F_{\nu}d^{2}}{\kappa_{\nu}B_{\nu}(T_{d})} \simeq 3.707 \bigg( \frac{F_{2 \rm mm}}{\rm mJy} \bigg) \bigg( \frac{d}{150\rm \ pc} \bigg)^{2} \, M_{\oplus},
\end{equation}

\noindent
where $F_{\nu}$ is the mm flux at 2.1 mm, $d$ is the source distance from the Gaia EDR3, $T_{d}$ is the assumed dust temperature, $B_{\nu}$ is the Planck function at $T_{d}$, and $\kappa_{\nu}$ is the dust grain opacity. To be able to compare with other measurements in the literature, we follow the same assumptions used by \citet{ansdell16,ansdell2018,tazzari2020a} in the Lupus surveys at 0.9, 1.3 and 3.0 mm, respectively. We adopted a power-law opacity of the form $\kappa_{\nu}$ = 10 ($\nu$/1000GHz)$^{\beta}$ cm$^{2}$g$^{-1}$, with a $\beta$ value of 1 \citep{beckwith90}, which yields $\kappa_{2.1\rm mm}$ = 1.46 cm$^{2}$g$^{-1}$, assuming isothermal dust with a temperature $T_{d}$ = 20 K, the median for Taurus disks \citep{andrews05}. 
Using equation (6) we found a dust mass of $M_{\rm dust}$ = 8.89 $M_{\oplus}$. This is consistent with other works in the literature stating that disk masses could often be underestimated by a large fraction by assuming optically thin emission \citep[e.g,][]{galvan-madrid2018}.  

This highlights the impact of assuming optically thin emission and not including scattering effects when estimating dust masses. Even though our optical depths are lower than 1 in all the data-sets, it seems that having only marginally optically thick regions may lead to an underestimation of the dust mass by using equation (6). Overall, it seems that the main factor contributing to the underestimation of dust masses in (sub)millimeter surveys is disregarding optical depth and scattering effects. This may be a solution to the mass budget problem for planet formation --the mass of solids in (sub)millimeter surveys of protoplanetary disks seems to be too low to explain the observed exoplanetary systems. 

\section{Discussion} 

The transition disk around the low-mass star Sz\,91 has a structure composed of mm-sized particles located in a well defined ring, a large dust depleted cavity and smaller dust particles inside this mm-cavity (as shown by polarimetric observations). Therefore, Sz\,91 represents a clear example of dust filtering and radially confined dust particles. 
We presented new ALMA band 4 observations of the dust ring and by combining the new data with previous ALMA 
observations in band 7 and band 6 we derived the spectral index, optical depth, and maximum grain size in the 
ring. We find the spectral indices as well as the maximum particle size to be nearly constant in the ring and the emission to be only marginally optically thick. 
In what follows we discuss the implications of these findings for the evolutionary status of the disk and compare Sz\,91 with the disk population in Lupus as well as with transition disks in other regions.

\subsection{Grain growth and possible planetesimal formation in the ring around Sz\,91}

Our results imply that dust particles in the Sz\,91 disk have grown to at least 0.61 mm size. Given the ring-like structure of this source it is now clear that dust is accumulating in a dust trap produced by the local pressure maxima sculpting the ring. ALMA observations are only sensitive to sub-mm particles, so cm particles may also be present in the disk. 
Therefore, understanding the disk around Sz\,91 requires modelling of particle growth, including fragmentation, and eventually 
planetesimal formation by the streaming instability, which has recently been reported to be a robust process in pressure bumps  \citep{carrera2021,guilera2020}. 

In this line of reasoning, \citet{stammler2019} explained the nearly constant optical depth of substructures 
in the DSHARP targets by including dust growth and fragmentation, and also planetesimal formation through the streaming instability in a one-dimensional simulation of dust evolution especially designed 
for the second dust ring of the protoplanetary disk around HD\,163296.
Even though HD\,163296 has a stellar luminosity of 17 $L_{\odot}$, significantly higher than Sz\,91,  
these simulations are particularly helpful for interpreting
our observational results as the ring in the model is located at a similar distance from the central star and is of a similar width as the mm-ring around Sz\,91. Moreover, since the model used in \citet{stammler2019} is composed of a disk with a gap at 83.5 au, the influence of the stellar irradiation will be more significant in the inner regions and should not affect the evolution of the second dust ring directly. 

The model by \citet{stammler2019} is largely based on \citet{birnstiel2010} 
but includes planetesimal formation when a dust to gas mass ratio of unity is reached in the disk midplane. 
While the equilibrium reached between grain growth and fragmentation is not affected by the inclusion of planetesimal formation, 
the optical depth is kept at values similar to those derived from observations of the DSHARP rings. 
The obtained peak optical depth in the simulations first increases with time until planetesimal formation due to the streaming instability removes mm particles from the midplane which leads to a decrease in optical depth. For ages ranging from 0.1 to 13 Myr the peak optical depth varies between 0.2 and 0.6 (see their Fig.\,2). The spectral index in the ring reaches nearly constant values of $3.0-3.5$
for ages between 1 and 13 Myr (their Fig.\,5). 

Sz\,91 is between 3 and 5 Myr old \citep{mauco2020}, the spectral index inside the ring is nearly constant with a value of 3.3-3.5, the peak optical depth we measure from our observations covers the range of 0.2-0.6 for the three observed bands if scattering is included and between 0.01 and 0.1 for the pure absorption case. 
The ring around Sz\,91 can therefore be added to the DSHARP sample of rings clustering in a close range of optical depth. 
In addition, both the spectral index as well as the optical depth perfectly match  the predictions of the model presented by \citet{stammler2019}. It appears therefore plausible that planetesimal formation is on-going in the ring around Sz\,91. 
While the agreement between predictions and observations seems robust, we advocate some caution as the model presented by \citet{stammler2019} does not include scattering effects in the determination of the optical depth, which is inconsistent with the values derived from observations that indicate that scattering plays a role (if the emission is not optically thin). Also, as noted by \citet{stammler2019} their model does not include 
the back reaction of dust particles onto the gas which can 
smear out concentrations \citep{taki2016,garate2019}. 

Alternatively, \citet{zhu2019} showed that optically thick scattering disks can also explain the peak optical depths observed in the DSHARP sample. They explained how the optically thin assumption may be incorrectly applied to an optically thick disk with reduced emission due to scattering. However, this scenario only applies to the inner disk within 50 au. Since our source consists of a dusty ring beyond 60 au, where the emission can be well fitted with a Gaussian profile along the radial direction, the scenario proposed by \citet{zhu2019} might not be applicable in our case. Furthermore, when the disk is large most of the dust mass is in the outer disk which is generally optically thin at 2.1 mm. If this is the case, then the disk mass obtained with equation (6), assuming optically thin emission, should underestimate the mass only by a factor of $\sim$2 \citep{zhu2019} compared to the mass obtained from our detailed analysis of the mm spectrum including scattering. This is similar to what we found in this work (within the uncertainties) considering that the emission is actually marginally optically thick. 

Additionally, the high value of the spectral index ($\alpha \sim$3.34) in the ring around Sz\,91 is also in line with the emission not being optically thick. Therefore, we can conclude that the dust emission at band 4 shown here can be used to properly characterize the dust particles in the ring which, according to our observational results, are composed of dust particles with maximum grain size of at least 0.61 mm and given the observed optical depth and the spectral index in the ring, on-going planetesimal formation is a likely scenario to explain these observational signatures.  

\subsection{Comparison with disks in Lupus}

\begin{figure}
	\includegraphics[width=\columnwidth]{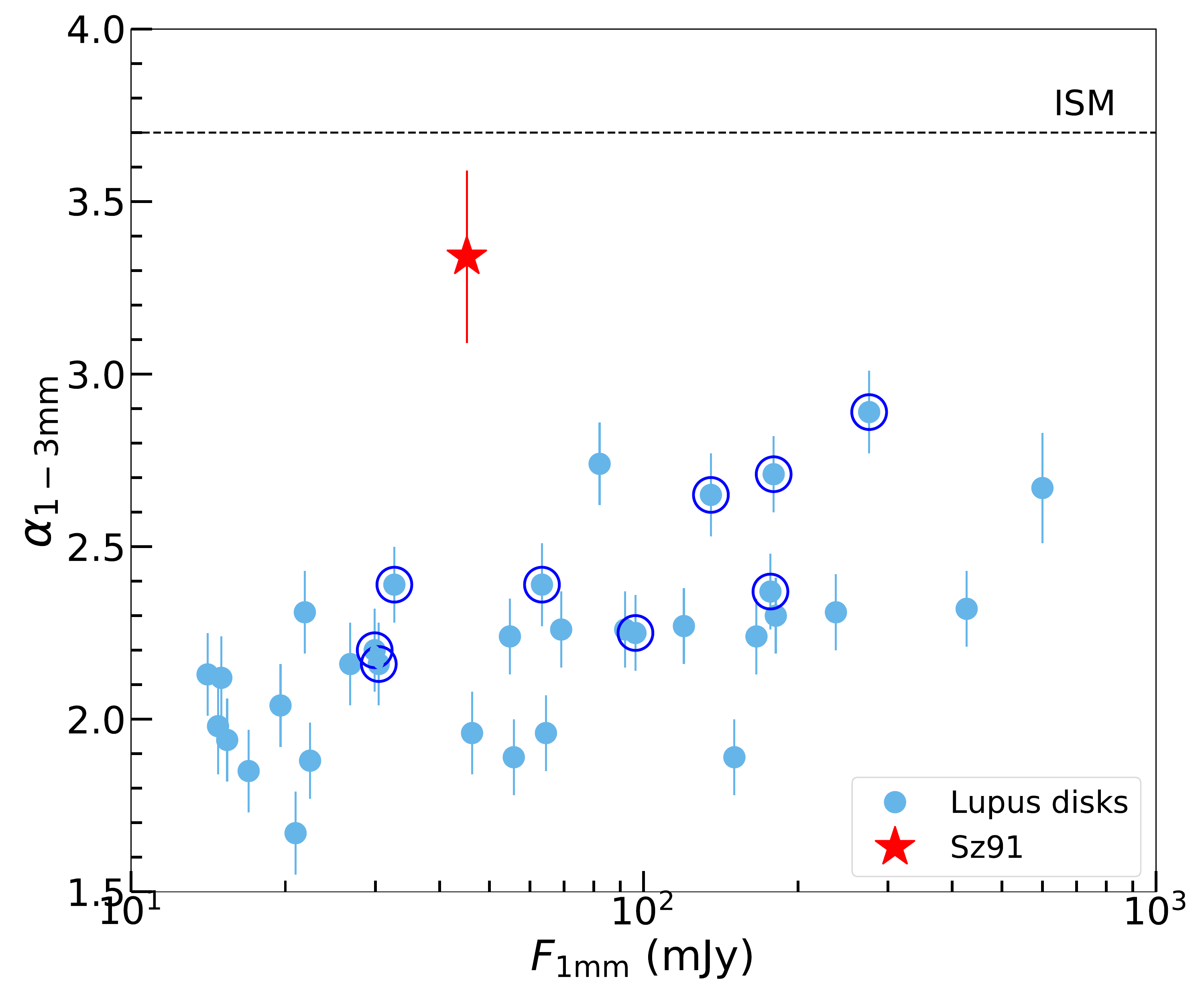}
    \caption{Spectral index between 1 and 3 mm as a function of integrated 1 mm flux for Lupus disks (blue filled circles) from \citet{tazzari2020a}. TDs are marked with an additional blue circle. The 1 mm flux is taken from \citet{ansdell2018}. The dashed line is the typical $\alpha_{1-3 \rm mm}$ of the optically thin emission of ISM dust. The red star indicates the position of Sz\,91 using our estimate of $\alpha$ from 1 mm to 2.1 mm. Sz\,91 appears as an outlier with the highest $\alpha$ value of 3.34 among the Lupus disk population.}
    \label{fig:alpha_Lupus}
\end{figure}

Thanks to the advent of ALMA, large surveys of disk populations have been conducted at millimeter wavelengths. Several authors have now focused their attention on tackling disk evolution through disk demographics. Two particular works have been done in the Lupus star-forming region aiming at characterizing grain growth \citep{tazzari2020a,ansdell16,ansdell2018}.
\citet{tazzari2020a} found spectral indices for the 35  brightest Class II disks of $\alpha_{1-3 \rm mm}$ $<$ 3 with a mean value of 2.23 (see Figure~\ref{fig:alpha_Lupus}). Note that, Sz\,91 was not observed in this work because of an erroneous observational setup. They found a tendency of larger values for transition disks ($\alpha_{\rm TD} \sim 2.5$, marked with an additional blue circle in Figure~\ref{fig:alpha_Lupus}), something also found at shorter wavelengths by \citet{ansdell2018} using fluxes from 0.9 to 1.33 mm. 
They interpreted this as evidence of grain growth (in the optically thin and Rayleigh-Jeans regime) with maximum grain sizes larger than 1 mm, for a range of reasonable dust composition and porosity. 

In the context of the Lupus disk population, Sz\,91 stands as the source with the highest spectral index. In Figure~\ref{fig:alpha_Lupus}, we compare the $\alpha_{1-3 \rm mm}$ reported in \citet{tazzari2020a} with that of Sz\,91 estimated here ($\alpha_{1-2.1 \rm mm}$, red star) as a function of their 1 mm fluxes. The spectral index of 3.34 plotted in the figure and found in sec~\ref{sec:spectral_indices} is totally consistent with the disk integrated spectral index estimated with the fluxes reported in sec~\ref{sec:obs}. At 220 mas resolution, for instance, we obtained a $\alpha_{1-2.1 \rm mm}$ = 3.31 $\pm$ 0.26. Consistent with earlier findings that TDs seem to have larger spectral indices than full disks, the spectral index of Sz91 is $\sim$1.5 times larger than the other disks in Lupus, and similar to the other transition disks in Lupus (identified in \citet{vandermarel2018}).

\citet{ansdell2018} mentioned that one of the reasons of the low $\alpha$, especially for brighter disks, may be attributed to larger optically thick regions in more massive disks, something also mentioned in \citet{galvan-madrid2018}. In this regard, we estimated the optical depths for the three data-sets used in this work. Figure~\ref{fig:tau_dist} 
shows that the emission in all three ALMA bands is optically thin if we consider only absorption (dashed lines), and is still lower than 1 in the more general case including scattering emission (solid lines). Given the large cavity around this source \citep[$R_{\rm cav} \sim$ 86 au;][]{francis2020}, where lower temperatures are expected for dust at such large radii, it is not surprising that dust emission is optically thin at mm wavelengths. The absence of optically thick regions in the disk around Sz\,91 may be the reason for its particularly high spectral index. 

Furthermore, as pointed by \citet{zhu2019}, if observations have measured that $\alpha$ is $\lesssim$ 2, it could be a strong indication that the disk is optically thick at those wavelengths and dust scattering plays an important role. They explained that values as low as 2 for the spectral index is not likely to happen in optically thin disks. If the disk is optically thick at the inner disk and optically thin at the outer disk, for instance, then the spectral index will be around 2 at the inner disk (which is optically thick) and suddenly change to 3-4 when $\tau <$ 1. This is simply because $\alpha$ in the optically thick and thin regimes are determined by different physical mechanisms. Therefore, the increase of $\alpha$ at the outer disk observed in some systems may be due to the whole disk becoming optically thin. In fact, this scenario is particularly suitable to the Lupus region since most of its disk population is composed of compact (small) disks \citep{ansdell2018, tazzari2020a} which are most likely optically thick at ALMA bands and hence will have lower spectral indices than more extended disks, as is the case of Sz\,91.

\subsection{Comparison with transition disks in other regions}

\begin{figure}
	\includegraphics[width=\columnwidth]{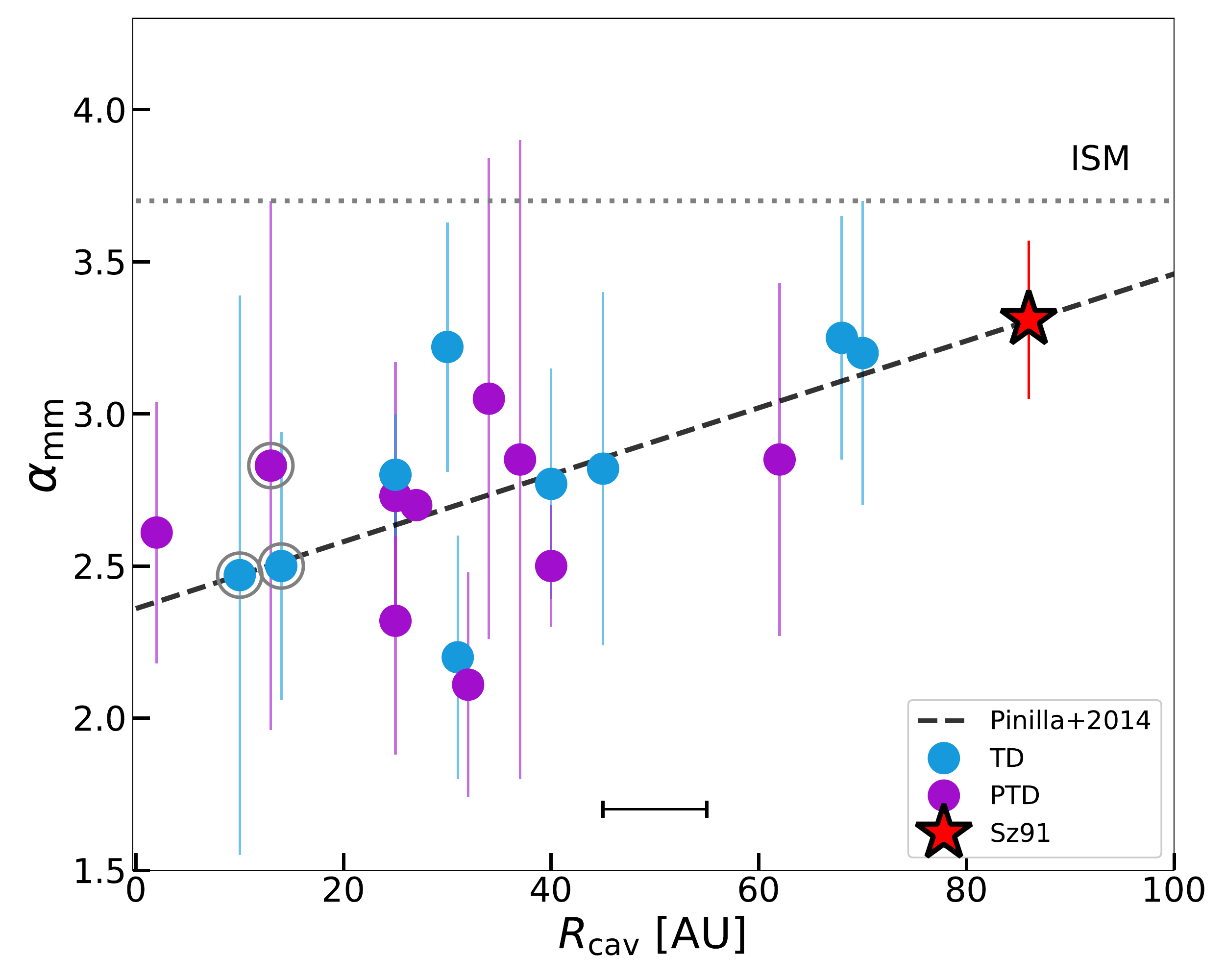}
    \caption{Integrated spectral index between $\sim$1 and $\sim$3 mm as a function of the cavity radius for the sample of TD in \citet{pinilla2014}. Sz\,91 is also plotted for comparison (red star). The dashed line shows the $\alpha_{\rm mm}$--$R_{\rm cav}$ relationship found in \citet{pinilla2014}. Objects with detected inner disks (pre-transition disk) are indicated as purple dots. The error bar at the bottom indicates the average error in the cavity radius. Sources with cavity sizes estimated through SED fitting are indicated with an additional gray circle.}
    \label{fig:alpha_TDs}
\end{figure}

To inquire if Sz\,91 also stands out when compared to other TDs in different star-forming regions, we plot in Figure~\ref{fig:alpha_TDs} the integrated spectral index from $\sim$1 mm to $\sim$3 mm, $\alpha_{\rm mm}$, as a function of cavity size for the sample of TDs reported in \citet{pinilla2014}. The values of $\alpha_{\rm mm}$ were taken from their tables 1 and 2, except for SR 24S and SR21 that were taken from the more recent works of \citet{pinilla2019} and \citet{pinilla2015c}, respectively.

The size of the cavity for each source, $R_{\rm cav}$, is taken from \citet{pinilla2014} except for those objects with updated sizes where we have used the most recent values: CS Cha, DM Tau, DoAr 44, GM Aur, MWC 758, T Cha, TWHya, UX Tau A, WSB 60 from \citet{francis2020}, J1604-2130, LkCa15, SR 21 from \citet{vandermarel2015}, SR 24 S from \citet{cieza2020}, and SZ Cha from \citet{ribas2016}. Sources with cavity sizes estimated through SED fitting are indicated with an additional gray circle. The error bar at the bottom shows the average error expected for the cavity size ($\pm$ 5 au). Finally, objects with detected inner disks in the \citet{francis2020} survey have been identified as pre-transition disks (PTD) in the figure. 

\citet{pinilla2014} found a correlation between the disk integrated spectral index and the cavity size in (pre)transition disks (dashed line, in Figure~\ref{fig:alpha_TDs}). They explained it because the mm emission is dominated by the dust at $R_{\rm cav}$ (i.e., in the pressure bump). Therefore, disks with wider cavities (pressure bumps located further out from the star) will have smaller $a_{\rm max}$. The smaller grains will then experience a lower radial drift making turbulent motions the main source of destructive collisions. In this case, the $a_{\rm max}$ is reached when the fragmentation velocity of the particles equals the turbulent relative velocity. At this point, $a_{\rm max}$ will also scale with the gas surface density of the disk which decreases with distance to the star.

Sz\,91 perfectly fits the $\alpha_{\rm mm}$--$R_{\rm cav}$ relationship of \citet{pinilla2014} which is consistent with our estimate of $a_{\rm max}$ = 0.61 mm, i.e., in the sub-mm range (see section~\ref{sec:radial_SED_fit}). In this regard, Sz\,91 follows the expected behavior of a TD with a large cavity. Since it hosts the largest cavity around a single T Tauri star (and the largest one in the sample plotted on the figure), it is expected to possess the highest spectral index which is, indeed, the case. This reinforces what we found in Figure~\ref{fig:alpha_Lupus}, where the significant difference between the spectral index of Sz\,91 and the disk population in Lupus is due to the latter being mostly composed of compact (small) disks.



\section{Conclusions}

In this paper we have presented new ALMA band 4 (2.1 mm) observations of the transition disk Sz\,91 combined with archival band 6 (1.3 mm) and band 7 (0.9 mm) data. 
The main results can be summarised as follows:

\begin{enumerate}

    \item We obtained 2.1 mm ALMA observations at $\sim$0.1{\arcsec} resolution and a sensitivity of 5.4 $\mu$Jy beam$^{-1}$. The continuum image shows a well resolved ring of dust peaking at $\sim$90 au from the central star. 
    
    \item By combining the new 2.1 mm observations with previous ALMA observations at 0.9 and 1.3 mm at 220 mas resolution, we derive the spectral index of the disk around Sz\,91 and find it to be constant throughout the ring with $\alpha \sim 3.34$ $\pm$ 0.26 and not showing clear azimuthal variations. Comparing this value with spectral indices reported in the literature for the disk population in Lupus we find that Sz\,91 exhibits the highest $\alpha$ of the region. Optically thick regions in the disks around non-resolved Lupus sources may account for their lower $\alpha$ values.  
    
    \item We estimated maximum grain sizes in the ring around Sz\,91 applying two different approaches. First, assuming optically thin emission and without including scattering effects: we find a slope of the absorption coefficient of $\beta_{\rm abs} = 1.34$ which requires grains with 1.0\,mm\,$<$\,$a_{\rm max}$\,$<$\,2.5\,mm. Second, in a more realistic approach including scattering emission and performing a radial analysis of the mm spectrum without putting any constraint in the optical depth at any wavelength (following \citealp{carrasco2019}): we find that the dust ring is composed of particles with a maximum grain size of $a_{\rm max}$ $ \sim$ 0.61\,mm. Scattering effects must be taken into account when characterizing the dust content in protoplanetary disks if one wants to avoid overestimating the maximum grain size.
    
    \item Sz\,91 perfectly fits the relationship found in \citet{pinilla2014} between the integrated spectral index and cavity size for transition disks. Given its large cavity ($\sim$86 au) it has the highest spectral index among the TD sample considered here. This is consistent with our $a_{\rm max}$ estimate lying in the sub-mm range and reinforces the fact that the significant difference between the spectral index of Sz\,91 and the Lupus population is due to the latter being mostly composed of small disks.
    
    \item The disk emission is marginally optically thick for the more general case (with scattering) with a peak optical depth between 0.2 and 0.6 decreasing down to $\tau_{\nu}\sim$0.01 for the longest wavelength (if scattering is excluded). These values for the disk around Sz\,91 are in the same range as those of the DSHARP targets.
    
    \item The total mass of solids that we obtain by integrating the expected surface density profile obtained with our multi-wavelength analysis, considering the effects of optical depth and self-scattering, is $M_{\rm dust}$ = 31.3$^{+6.3}_{-10.6}$ $M_{\oplus}$. Lower dust masses, by at least a factor of 2, are found if one assumes the emission to be optically thin.    
\end{enumerate}

We interpret these findings as evidence of on-going grain growth produced by the trapping of dust in a pressure bump.  
The nearly constant spectral index and range of optical depth values found in the ring around Sz\,91 agree very well with the predictions of planetesimal formation by \citet{stammler2019}. Given the self-regulating nature of this process, by stabilizing the dust-to-gas mass ratio in the midplane added to the steady state reached between particle growth and fragmentation, after 1 Myr the spectral index in the ring has reached its minimum value and enough mm particles have been converted to planetesimals as to constrain the optical depths to the observed values. Sz\,91, therefore, represents a plausible case of possible planetesimal formation in a transition disk. Future multi-wavelength observations resolving the disk ring radially could provide crucial additional constraints on models of planetesimal and planet formation. 

\acknowledgments

The authors thank the anonymous referee for a careful reading of our
manuscript and for the helpful comments that improved the content and presentation of this work.

KM acknowledges financial
support from CONICYT-FONDECYT project no. 3190859. KM, JO, MRS, AB, and CC acknowledge support by ANID, -- Millennium Science Initiative Program -- NCN19\_171.
This work was supported by UNAM DGAPA-PAPIIT grants IN108218, IN111421 and IG101321 and CONACyT Ciencia de Frontera grant number 86372.
MRS acknowledges support from Fondecyt (grant 1181404).  
AB acknowledges support from FONDECYT Regular 1190748.
JO acknowledges support from the Universidad de Valpara\'iso, and from FONDECYT Regular 1180395.
CC acknowledges support from DGI-UNAB project DI-11-19/R.
This work has made use of data from the European Space Agency (ESA) mission
A.S. acknowledges support from ANID/CONICYT Programa de Astronom\'ia Fondo ALMA-CONICYT  2018 31180052. 
{\it Gaia} (\url{https://www.cosmos.esa.int/gaia}), processed by the {\it Gaia}
Data Processing and Analysis Consortium (DPAC,
\url{https://www.cosmos.esa.int/web/gaia/dpac/consortium}). Funding for the DPAC
has been provided by national institutions, in particular the institutions
participating in the {\it Gaia} Multilateral Agreement.
This paper makes use of the following ALMA data: \\

ADS/JAO.ALMA\#2018.1.01020.S, 

ADS/JAO.ALMA\#2012.1.00761.S and

ADS/JAO.ALMA\#2015.1.01301.S. ALMA is a partnership of ESO (representing its member states), NSF (USA) and NINS (Japan), together with NRC (Canada), MOST and ASIAA (Taiwan), and KASI (Republic of Korea), in cooperation with the Republic of Chile. The Joint ALMA Observatory is operated by ESO, AUI/NRAO and NAOJ. In addition, publications from NA authors must include the standard NRAO acknowledgement: The National Radio Astronomy Observatory is a facility of the National Science Foundation operated under cooperative agreement by Associated Universities, Inc.

%

\vspace{3mm}
\facilities{ALMA}


\software{Astropy \citep{astropy13},  
          CASA \citep{McMullin2007}, 
          Matplotlib \citep{matplotlib05},
          Numpy \citep{numpy2011}
          }



\vspace{3mm}

\appendix

\section{SED fitting using different power-law indices for the particle size distribution}

\begin{figure}
	\includegraphics[width=0.5\textwidth]{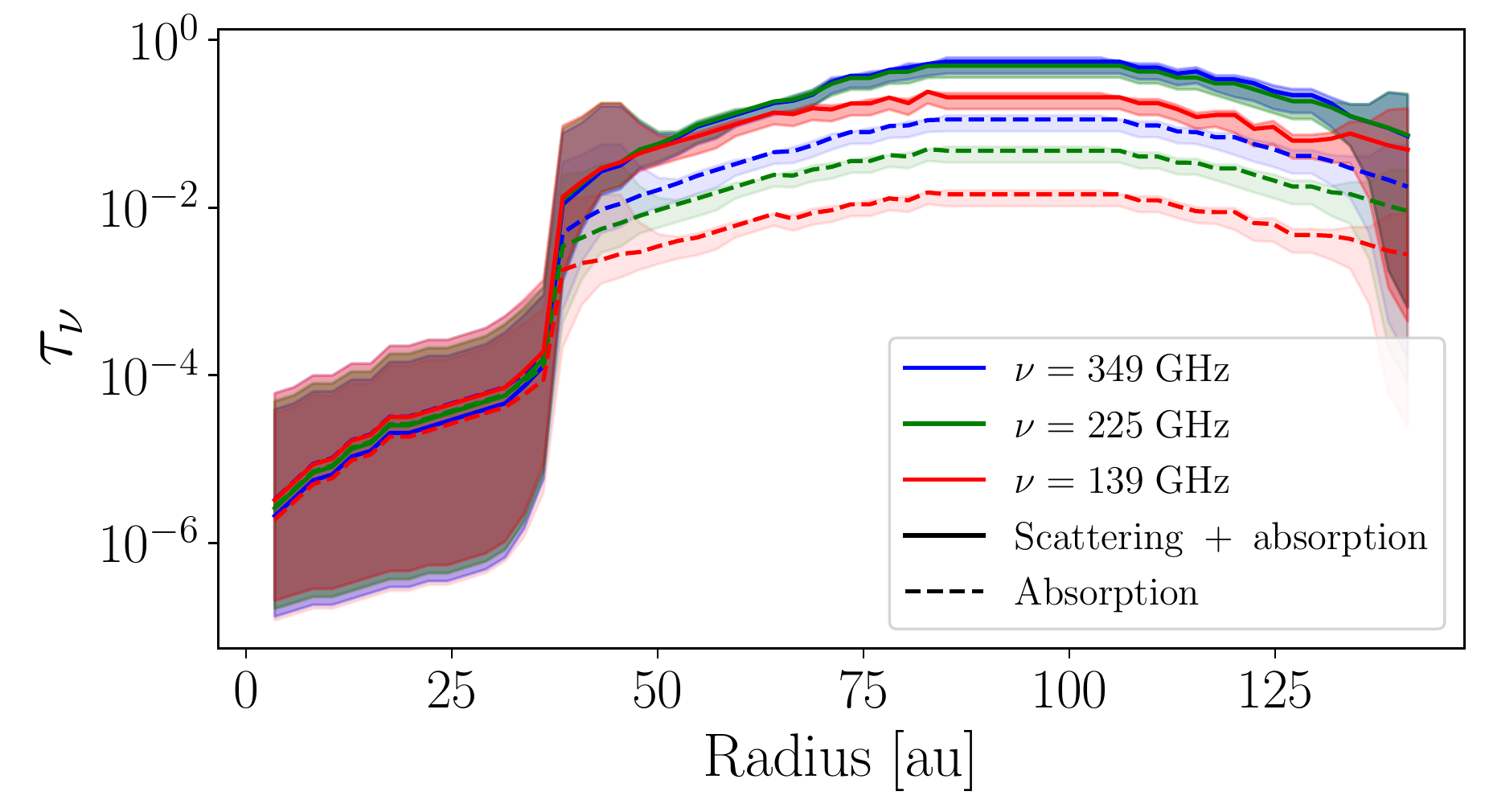}
	\includegraphics[width=0.5\textwidth]{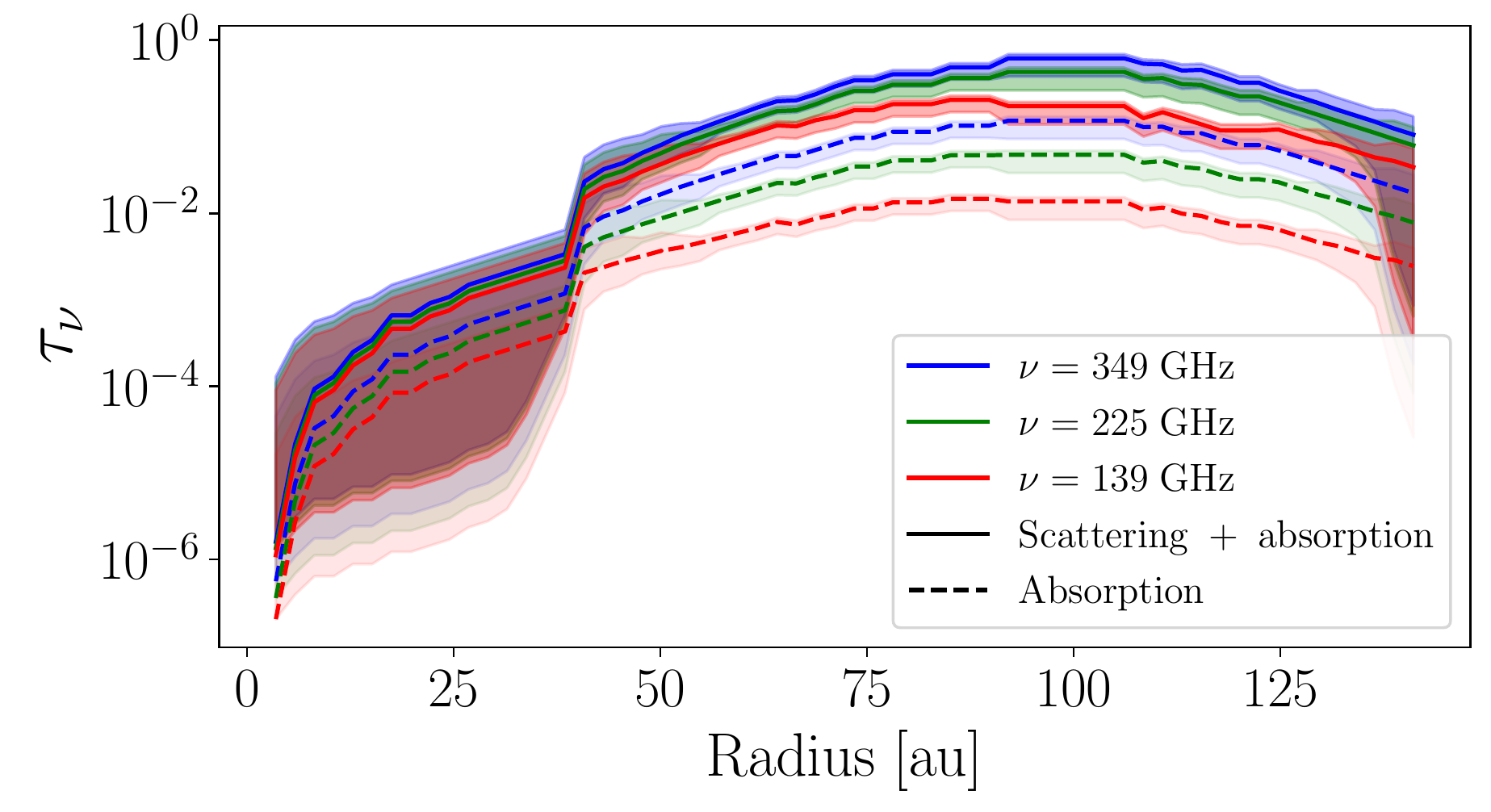}

    \caption{Optical depth profiles for the ALMA band 4 (red), band 6 (green), and band 7 (blue) observations estimated using a slope for the particle size distribution of $q = 2.5$ (left) and $q = 3.5$ (right). Dashed lines indicate pure absorption cases while solid lines show optical depths for cases also including scattering. 2 $\sigma$ level uncertainties are shown as shaded regions.}
    \label{fig:tau_dist_difq}
\end{figure}

\begin{figure}
	\includegraphics[width=\textwidth]{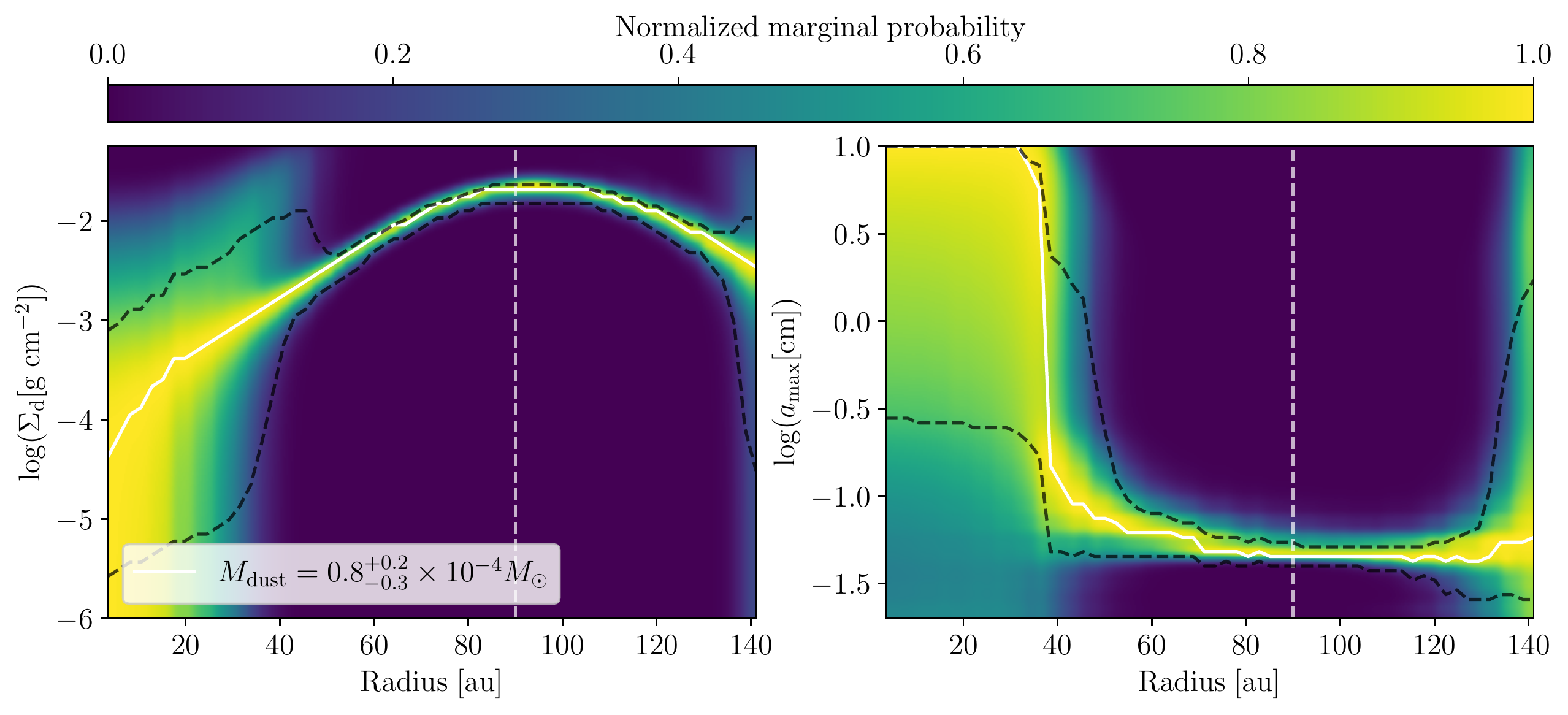}
    \caption{Radial fitting of the mm spectrum (see Section~\ref{sec: amax and tau} for details) using a power-law index of the particle size distribution of $q = 2.5$. Probability distributions of the surface density (left) and maximum grain size (right) are shown as a function of radius, where the colorbar shows the marginal probability. White solid lines indicate the best fit. Black dashed lines represent 2 $\sigma$ level uncertainties. Vertical lines indicate the location where the dust emission peaks (90 au). The total dust mass in the ring is 27$^{+7}_{-10}$ $M_{\oplus}$ and the maximum grain size $\sim$ 0.56 mm.}
    \label{fig:amax_dist_q2.5}
\end{figure}

The optical depth profiles obtained for the ring around Sz\,91 using a power-law index for the particle size distribution of $q = 2.5$ are shown in Figure~\ref{fig:tau_dist_difq} (left panel). Red, green and blue lines indicate the ALMA band 4, 6, and 7 optical depths, and dashed lines represent the pure absorption case while solid lines trace cases also considering scattering. The resulting grain sizes and dust surface densities, on the other hand, are shown in Figure~\ref{fig:amax_dist_q2.5}. We found a $a_{\rm max}$ = 0.56 mm and a $M_{\rm dust}$ = 27$^{+7}_{-10}$ $M_{\oplus}$.

\begin{figure}
	\includegraphics[width=\textwidth]{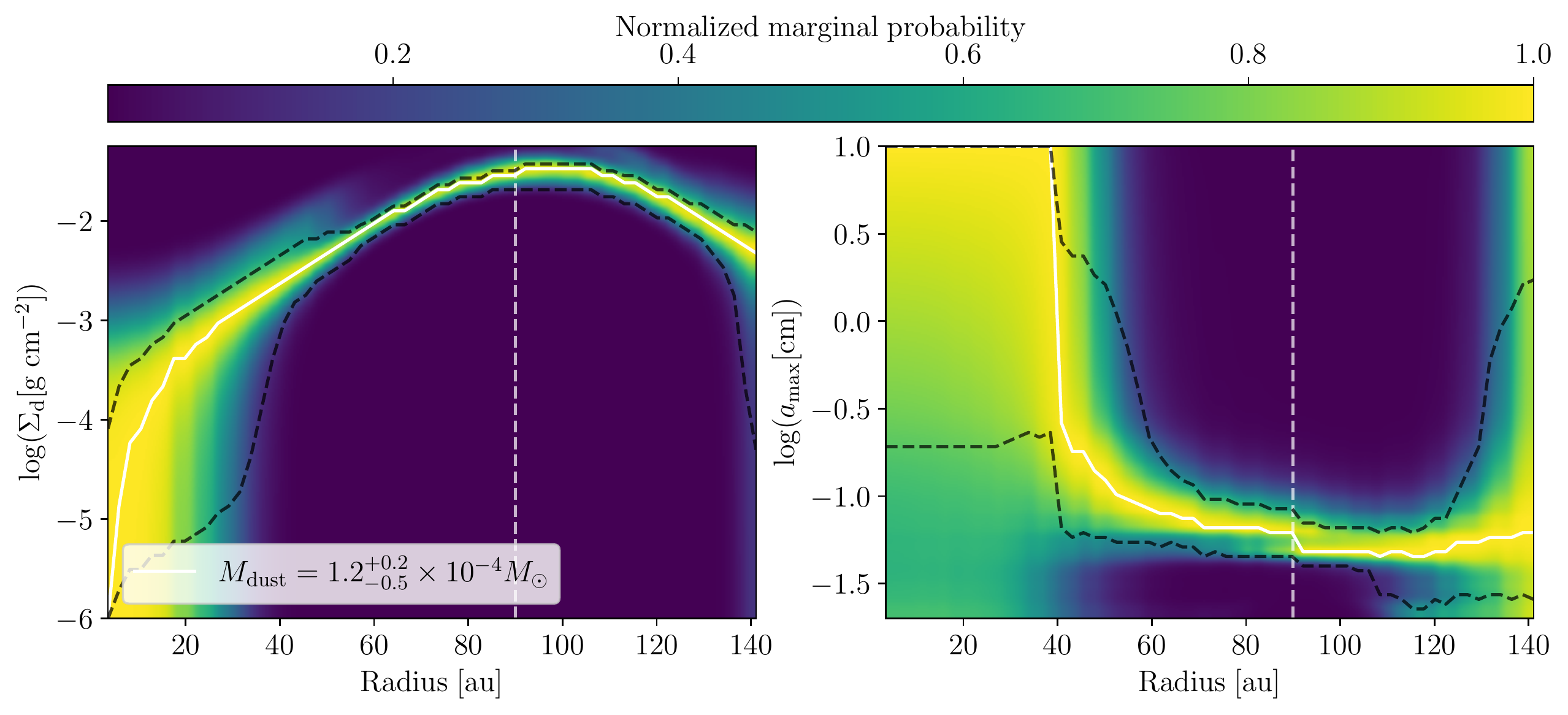}
    \caption{Radial fitting of the mm spectrum (see Section~\ref{sec: amax and tau} for details) using a power-law index of the particle size distribution of $q = 3.5$. Probability distributions of the surface density (left) and maximum grain size (right) are shown as a function of radius, where the colorbar shows the marginal probability. White solid lines indicate the best fit. Black dashed lines represent 2 $\sigma$ level uncertainties. Vertical lines indicate the location where the dust emission peaks (90 au). The total dust mass in the ring is 40$^{+7}_{-17}$ $M_{\oplus}$ and the maximum grain size $\sim$ 0.65 mm.}
    \label{fig:amax_dist_q3.5}
\end{figure}

These results are similar to the canonical value of $q = 3.5$ shown in Figure~\ref{fig:tau_dist_difq} (right panel) and Figure~\ref{fig:amax_dist_q3.5}, where symbols and colors are the same as in Figure~\ref{fig:amax_dist_q2.5}. The main difference is that the maximum grain size is less constrained in the canonical case with an average value of 0.65 mm and the dust mass is 40$^{+7}_{-17}$ $M_{\oplus}$, 1.5 times higher than for the case of $q=2.5$.  

\section{Line Imaging}

Even though the main goal in this work is based on continuum observations, we centered 1 of our 4 spws at the rest frequency of the Carbon monosulfide (CS v=0, 3--2) line at 146.96 GHz aiming for a possible detection. 

\begin{figure}
	\includegraphics[width=1.0\textwidth]{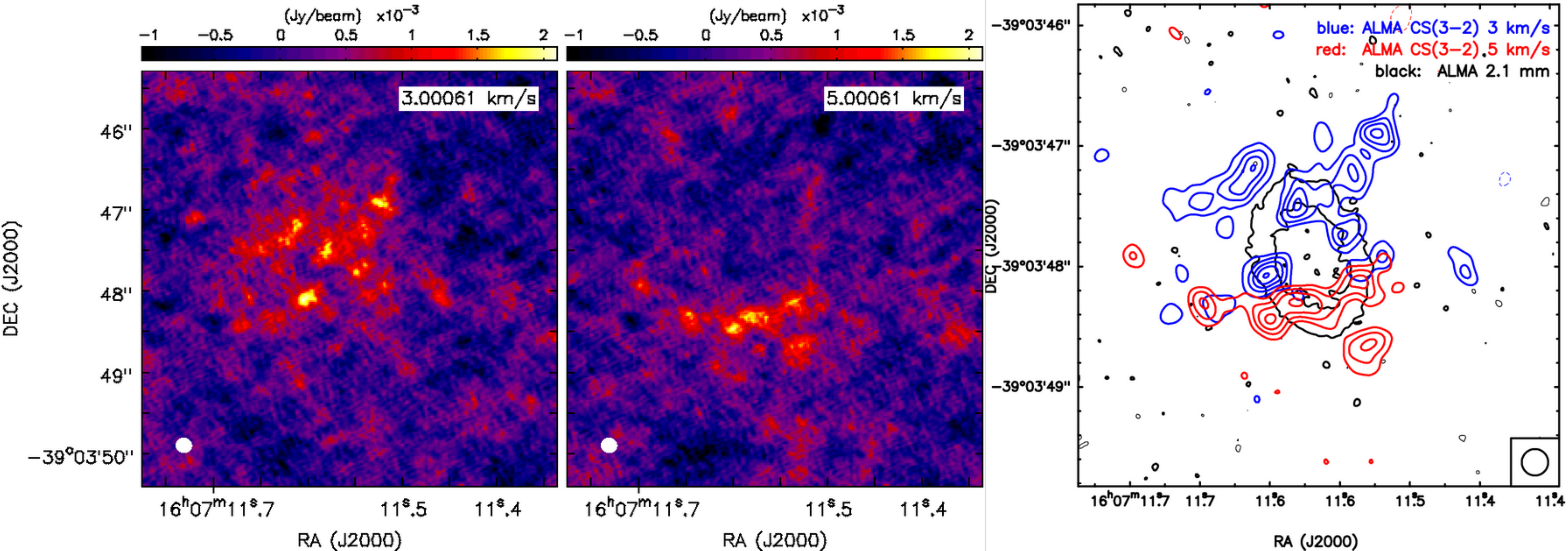}
    \caption{\textit{Left:} CS v = 0, (3--2) line continuum subtracted CLEAN image. We detect the line at the velocity range from $\sim$3 km s$^{-1}$ to $\sim$5 km s$^{-1}$. The emission peaks at 3 km s$^{-1}$ with and intensity peak of 2.11 mJy beam$^{-1}$ (6.5 $\sigma$). The beam size of 197\,$\times$\,180 mas is shown as a filled circle at the bottom left of each panel. \textit{Right}: CS v = 0, (3--2) emission of the blue and red channels of the cube after smoothing to a final beam of 0.22{\arcsec} (shown at the bottom right). The blue and red contours correspond to 3, 4, 5, 6, and 7 times the rms of the smoothed cube ($\sim$0.33 mJy/beam). The black contours correspond to the 3$\sigma$ contour of the 2.1 mm continuum image. }
    \label{fig:CSLine_ima}
\end{figure}

As shown in Figure~\ref{fig:CSLine_ima}, we detected the CS(3--2) line at velocities from $\sim$3 km s$^{-1}$ to $\sim$5 km s$^{-1}$. The RMS of the CLEAN image is 0.33 mJy beam$^{-1}$. The emission peaks at $\sim$3 km s$^{-1}$ with an intensity peak of 2.11 mJy beam$^{-1}$ (6.5 $\sigma$). 
The right panel shows the CS(3--2) emission of the blue and red channels of the cube after smoothing to a final beam of 0.22{\arcsec} (shown at the bottom right). The blue and red contours correspond to 3, 4, 5, 6, and 7 times the rms of the smoothed cube ($\sim$0.33 mJy/beam). It is worth noting that there are at least 5 peaks of CS emission that spatially match very well with the dusty ring. Out of these 5 peaks of gas emission, 2 peaks are detected at 5$\sigma$, 2 peaks are detected at 6$\sigma$ and one peak is detected at 7$\sigma$. Therefore, we are certainly detecting CS(3--2) gas associated with the dusty ring. In addition, there are other peaks further away that could be related to accretion streams or be simply part of the Keplerian pattern already shown by \citet{tsukagoshi2019} at CO(3--2) and HCO$^+$(4--3).



\begin{figure*}
	\includegraphics[width=\textwidth]{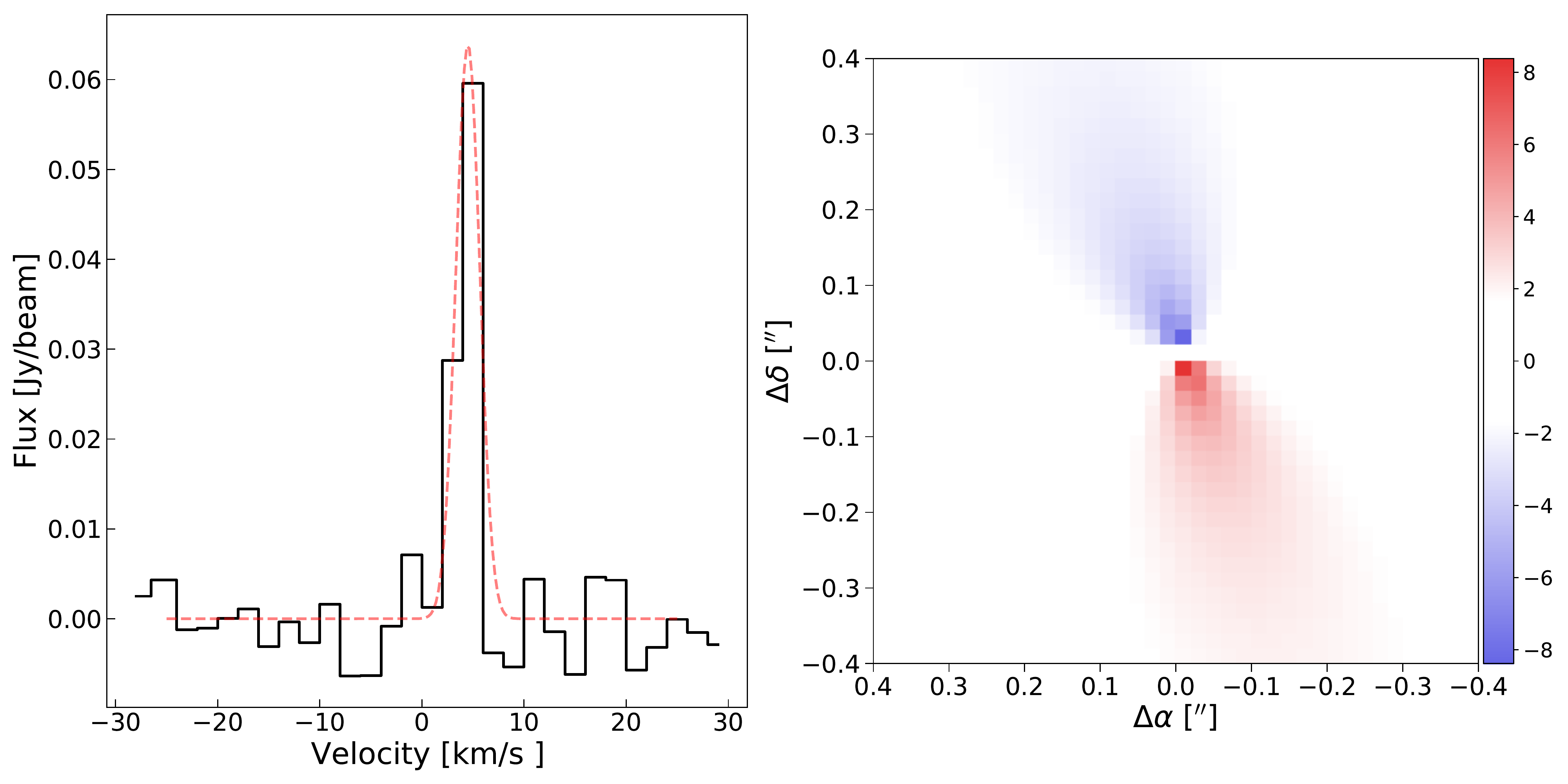}
    \caption{CS Line Spatial Filtering results. \textit{Left}: shifted spectrum. The line is clearly detected at 4.5 $\pm$ 0.1 km s$^{-1}$ given by the Gaussian fit (light-red dashed line). \textit{Right}: disk model. The color coding corresponds to the estimated Keplerian velocity for each pixel.}
    \label{fig:CSLine_spatial_filtering}
\end{figure*}

In order to boost the signal of the CS line (3--2) at 146.96 GHz we used the spectral and spatial filtering technique \citep{matra2017}. The idea is to ``correct" the velocity of each pixel in the data cube to account for the keplerian rotation of the disk. To this end, for each pixel, we compute the Keplerian velocity, accounting for the inclination and position angle of the disk obtained from the ALMA band 7 (0.9 mm) observations of \citet{tsukagoshi2019}. We assume a stellar mass of 0.58 $M_{\odot}$ \citep{mauco2020}, and we test for both direction of rotation. Then, for each pixel of the data cube, we shift the 1D spectrum by the opposite of the estimated velocity. Afterwards, for each velocity frame, we sum the total flux in a circular aperture of 1.5\arcsec\,size. If a line is present in the observations, its signal will be significantly increased, as it should all be shifted into a single (ideally) velocity frame. The two main caveats with this approach is that we assume the disk to be vertically flat, and that the gas is rotating at Keplerian velocity (while it may be rotating at sub-Keplerian velocity as it is self-supported by its own pressure).  Figure\,\ref{fig:CSLine_spatial_filtering} shows the result of the analysis. The left panel shows the shifted spectrum, and a line is clearly detected at 4.5 $\pm$ 0.1\,km s$^{-1}$, as shown by the Gaussian fit (light-red dashed line). This value is consistent with the radial velocity of the star reported in \citet[][3.4 $\pm$ 0.2 km s$^{-1}$]{canovas2015a} within a factor less than one. The right panel shows the disk model where the color coding corresponds to the estimated Keplerian velocity for each pixel. 

The CS line traces dense gas and it is now commonly observed in protoplanetary disks \citep{legal2019,teague2018,phuong2018,dutrey2017,guilloteau2016,fuente2010}. However, it is mostly observed in one of its higher transitions (5--4, 6--5). Here we report the less common lower (3--2) transition that can be used in combination with future detections of CS at higher transitions in order to estimate the excitation temperature of the gas emitting the line.

\bibliographystyle{aasjournal}



\end{document}